\documentclass[a4paper,12pt]{article}
\usepackage{jheppub,esint,shuffle,psfrag}
\usepackage[utf8]{inputenc}

\usepackage{epsfig,amssymb,amsmath,psfrag,subfigure,rotate,color,wasysym,xcolor}
\usepackage[bbgreekl]{mathbbol}

\allowdisplaybreaks 
\newcommand{\insertfig}[2]{\includegraphics[width=#1cm]{#2}}

\DeclareSymbolFontAlphabet{\mathbbm}{bbold}
\DeclareSymbolFontAlphabet{\mathbb}{AMSb}%

\def\XXint#1#2#3{{\setbox0=\hbox{$#1{#2#3}{\int}$ }
\vcenter{\hbox{$#2#3$ }}\kern-.6\wd0}}

\newcommand*\xbar[1]{%
  \hbox{%
    \vbox{%
      \hrule height 0.5pt 
      \kern0.3ex
      \hbox{%
        \kern-0.2em
        \ensuremath{#1}%
        \kern-0.1em
      }%
    }%
  }%
} 
\def \be  {\begin{equation}}
\def \ee  {\end{equation}}
\def \ba  {\begin{eqnarray}}
\def \ea  {\end{eqnarray}}
\def \baa {\begin{eqnarray*}}
\def \eaa {\end{eqnarray*}}
\newcommand{\ep}{\varepsilon}
\def \lab #1 {\label{#1}}

\newcommand\re[1]{(\ref{#1})}
\def\d{\hbox{{d}\kern-.20em\hbox{l}}}

\def \matrix #1 {\left(\begin{array}{cc} #1 \end{array}\right)}

\def \tr {\mathop{\rm tr}\nolimits}

\def \e  {\mathop{\rm e}\nolimits}

\newcommand \vev [1] {\langle{#1}\rangle}

\newcommand \ket [1] {|{#1}\rangle}
\newcommand \bra [1] {\langle {#1}|}
\newcommand{\bit}[1]{\mbox{\boldmath$#1$}}
\def\1{\hbox{{1}\kern-.25em\hbox{l}}}

\newcommand{\ft}[2]{{\textstyle\frac{#1}{#2}}}

\makeatletter

\input pdf-trans
\newbox\qbox
\def\usecolor#1{\csname\string\color@#1\endcsname\space}
\newcommand\bordercolor[1]{\colsplit{1}{#1}}
\newcommand\fillcolor[1]{\colsplit{0}{#1}}
\newcommand\outline[1]{\leavevmode%
  \def\maltext{#1}%
  \setbox\qbox=\hbox{\maltext}%
  \boxgs{Q q 2 Tr \thickness\space w \fillcol\space \bordercol\space}{}%
  \copy\qbox%
}
\makeatother
\newcommand\colsplit[2]{\colorlet{tmpcolor}{#2}\edef\tmp{\usecolor{tmpcolor}}%
  \def\tmpB{}\expandafter\colsplithelp\tmp\relax%
  \ifnum0=#1\relax\edef\fillcol{\tmpB}\else\edef\bordercol{\tmpC}\fi}
\def\colsplithelp#1#2 #3\relax{%
  \edef\tmpB{\tmpB#1#2 }%
  \ifnum `#1>`9\relax\def\tmpC{#3}\else\colsplithelp#3\relax\fi
}
\bordercolor{black}
\fillcolor{white}
\def\thickness{.3}

\def\1{\mathbbm{1}}

\title{Pinching Sudakov}
 
\author[a]{A.V.~Belitsky,}
\author[b]{L.V.~Bork,}
\author[c]{V.A. Smirnov}
\affiliation[a] {Department of Physics, Arizona State University,  Tempe, AZ 85287-1504, USA}  
\affiliation[b]{Institute for Theoretical and Experimental Physics, 117218 Moscow, Russia}
\affiliation[]{The Center for Fundamental and Applied Research, 127030 Moscow, Russia}
\affiliation[c]{Skobeltsyn Institute of Nuclear Physics, Moscow State University 119992 Moscow, Russia}
  
 \abstract
{In this paper, we discuss the factorization of the Sudakov form factor on the Coulomb branch of maximally supersymmetric Yang-Mills
theory in the near mass-shell limit. We unravel all pinch singularities of this observable making use of the Method of Regions. We find
their operator content in terms of matrix elements of Wilson lines on semi-infinite and finite intervals for the jet and ultrasoft functions,
respectively. However, naive factorization into these incoherent momentum components is broken at two-loop order by effects subleading
in the parameter of dimensional regularization. To save the day, we perform an appropriate twisting of the functions involved as well
as simultaneous finite scheme transformation of the 't Hooft coupling. Infrared physics of twisted jet and ultrasoft functions is governed 
by the octagon anomalous dimension, while the untwisted ultrasoft function possesses infrared evolution driven by an anomalous 
dimension different from the ubiquitous cusp.}

\begin{document}

\maketitle
\flushbottom
\setcounter{footnote} 0

\section{Introduction}

Quantum-mechanical independence of physics happening at different space-time or momentum scales is at the heart of factorization 
theorems which form, in turn, the foundation for quantitative applications of QCD to particle phenomenology. Factorization of a full 
amplitude into its incoherence components intrinsically introduces an arbitrary scale into the problem. Then the statement of independence 
of the former on the latter translates into the renormalization group equations. These equations allow one to perform an effective 
resummation of large corrections to all orders of perturbation theory. They stem from logarithms of ratios of the aforementioned 
different distance scales. The only first-principle ingredients required to accomplish this goal are the anomalous dimensions.

Some time ago \cite{Belitsky:2022itf}, we proposed an exact formula for the near mass-shell Sudakov form factor in the maximally
supersymmetric Yang-Mills there, aka $\mathcal{N} = 4$ sYM, based on a direct three-loop calculation \cite{Belitsky:2023ssv} as 
well as complimentary information from four-point correlation functions of infinitely heavy BPS operators \cite{Belitsky:2019fan}. 
The formula admits a stunningly simple form both in the 't Hooft coupling and in the kinematical $m^2 \equiv P^2/q^2$ invariant 
involved as $m \to 0$,
\begin{align}
\label{ExactOffSud}
\log \mathcal{F}_2 = - \ft12 \Gamma_{\rm oct} (g) \log^2 m^2 -  D (g)
\, ,
\end{align}
with
\begin{align}
\Gamma_{\rm oct} (g)
= \frac{2}{\pi^2} \log \cosh(2\pi g)
\, , \qquad
D (g) = \frac{1}{4} \log\frac{\sinh (4 \pi g)}{4 \pi g}
\, ,
\end{align}
where the coefficient accompanying the double logarithm is known as the octagon anomalous dimension. This result appears to clash with
folklore wisdom that all infrared-sensitive physics in gauge theories is driven by the so-called cusp anomalous dimension. The one-loop result
displays a well-known factor-of-two difference between the on- and off-shell kinematics \cite{Mueller:1979ih,Korchemsky:1988hd}, see also 
\cite{Forte:2020fbc} for a nice historical exposition\footnote{We would like to thank Lorenzo Magnea for drawing our attention to this paper.}. A 
factorization formula for this observable within QCD was suggested almost four decades ago \cite{Korchemsky:1988hd}. However, it refers 
only to the cusp and no trace of the octagon anomalous dimension can be found there. So how do we reconcile these? With the unfactorized 
form factor displaying such a simple form, does its separation into soft and collinear degrees of freedom obscure its simplicity such that it is 
gone at the end? Does this suggest that factorization is violated?

Most of the time, factorization is discussed in generic terms based on Landau equations \cite{Landau:1959fi} via Coleman-Norton adaptation 
\cite{Coleman:1965xm} and deducing corresponding pinch surfaces where singularities of various momentum modes reside. However, this is 
rarely done with support from explicit (multi) loop calculations. Here we go in the opposite direction. We have an exact answer for the form 
factor and diagrammatic representation for the lowest few orders of perturbation theory. The latter is encoded in a small set of scalar integrals 
corresponding to sums of the original Feynman graphs making use of Passarino-Veltman reductions. In fact, in the case at hand, these 
expressions were fast-tracked to their concise form making use of unitarity-based techniques. So our starting point will not be a 
diagram-by-diagram basis but rather a small set of scalar integrals. 

Our consideration in this paper will be based on the Method of Regions (MofR) \cite{Beneke:1997zp}, see Ref.\ \cite{Smirnov:2021dkb} 
for a concise review. This formalism, though carries on a label of experimental mathematics, had proven itself in the past to provide 
correct results for asymptotic expansions of various observables,
cross sections, and scattering amplitudes alike. MofR will provide us with a bare, i.e., before renormalization, form of the factorization 
formula we are looking for. The formalism provides an exact mapping of regions to pinch surfaces encoded in Landau equations. 
Our attachment to the method will immediately imply however that our factorization formula will be oblivious to certain incoherent 
components of the factorized result. Namely, since dimensional regularization/reduction is indispensable for a robust and proper 
treatment of different regions of the phase space, zero-bin effects \cite{Manohar:2006nz,Dixon:2008gr} will be invisible to us. This 
is a mere consequence of the fact that scaleless integrals identically vanish within it. However, the upside is that MofR gives a precise 
definition to contributing Feynman integrals in various momentum regions inherit the intrinsic property of the method: all loop-momentum 
integrals are unrestricted and cover the entire Minkowski space rather than a subspace with proper momentum scalings as practiced in 
certain approaches to factorization. This feature will offer us a way to uncover their operator definitions in terms of local fields and 
extended objects like Wilson lines. These can then be computed from the conventional Feynman diagram technique and compared with 
results from MofR.

Let us point out an advantage of applying MofR to the off-shell Sudakov form factors, compared to its on-shell counterpart. Factorization 
theorem for the latter was repeatedly discussed in the literature over the past forty years, see \cite{Agarwal:2021ais} for a thorough recent 
review, and are summarized in the following formula \cite{Dixon:2008gr}
\begin{align}
\label{OnShellSudFact}
\mathcal{F}_2^{\rm on-shell} =  H (J_1/J^{\rm eik}_1) (J_2/J^{\rm eik}_2) S
\, ,
\end{align}
where $J_i$ and $J^{\rm eik}_i$ are the jet functions \cite{Collins:1989bt} and their eikonal versions \cite{Dixon:2008gr}, 
respectively, $S$ is the soft function determined by the vacuum expectation value of semi-infinite Wilson lines meeting 
at a cusp \cite{Korchemsky:1987wg}, and, last but not least, $H$ is the hard matching coefficient. With the use of conventional 
dimensional regularization to tame both ultraviolet and infrared divergences, all quantum corrections to on-shell case matrix 
elements of operators built solely from Wilson lines vanish identically since they are given in terms of scaleless momentum 
integrals. On the one hand, it 
manifests the equivalence of infrared and ultraviolet effects in these functions, which was one of the reasons for using the 
renormalization group of Wilson lines for studies of the infrared physics of amplitudes \cite{Korchemsky:1987wg}. However, it 
then requires a clean separation between the two, and therefore, there is a necessity to rely on a regularization for infrared physics 
that is different from dimensional. In this manner, the eikonal functions become given by counterterms \cite{Gardi:2009qi}. On 
the other hand, if one entirely relies on dimensional regularization, which does the job perfectly, then the soft function and the 
eikonal jets are simply one. With this perfectly valid point of view, the soft effects migrate elsewhere in the above factorization 
formula, as was done multiple times in the past \cite{DelDuca:1990gz,Bonocore:2014wua}. However, in this
manner MofR ``throws the baby out with the tubwater" as one of the important ingredients of infrared physics becomes a collateral
of the formalism. None of these is the case for the off-shell Sudakov since the external virtuality introduces an intrinsic scale in
majority of matrix elements involving Wilson lines and so we can clearly identify all momentum components in the factorization formula
analogous to \re{OnShellSudFact} for $\mathcal{F}_2$ except for the eikonal jets (zero-bin subtractions). However, This will suffice 
for the goals we set up in this work.

Our subsequent presentation is organized as follows. In the next section, we recall the definition of the Sudakov form factor and
the basis scalar integrals defining its two-loop form. Next, in Sect.\ \ref{MofRfactor}, we review the geometric approach to MofR 
and apply it to the one-loop Sudakov form factor in the near mass-shell limit. We use these one-loop results to propose an operator
definition for the jet and ultrasoft functions which define leading pinched regions as $m\to 0$. In Sect.\ \ref{JetSect}, we evaluate the 
jet function to one loop order using its definition as an amputated off-shell Green function of a semi-infinite Wilson line on the light 
cone. In Sect.\ \ref{SectUSdef}, we define the ultrasoft function and, then, calculate it in Sect.\ \ref{US1looSect} and \ref{US2looSect} 
to one- and two-loop order, respectively. In Sect.\ \ref{AllOrderFactSect}, we conjecture a factorized form of the near mass-shell 
Sudakov form factor akin to Eq.\ \re{OnShellSudFact} in the on-shell case. Then, we turn to its verification at two-loop order making 
use of MofR in Sect.\ \ref{MofR2loopTestSect}. We verify there the equivalence of the ultrasoft region stemming from MofR to the 
one from the operator definition of the function in question. However, we encounter difficulties with the suggested naive factorization 
formula. Namely, while the effects of $O (\varepsilon)$ are irrelevant for the complete off-shell factor: the form factor is finite for 
$P_i^2 \neq 0$ and $\varepsilon$ can be set to zero in the sum, its individual factorized components are highly sensitive to these 
effects in order to enforce an agreement with the full expression. As we will demonstrate in Sect.\ \ref{TwistSect}, a slight mismatch 
between the product of the one-loop collinear and soft contributions and their two-loop counterpart forces us to `twist' the definitions 
of individual momentum components and also to perform a finite scheme transformation of the 't Hooft coupling. These steps change 
the form of the infrared evolution equations for the ultrasoft-collinear functions involved. Finally, we conclude. Two appendices contain 
supplementary material needed for a better understanding of the main body of the paper.

\section{Off-shell Sudakov form factor}
\label{OffShellSudakSect}

\begin{figure}[t]
\begin{center}
\mbox{
\begin{picture}(0,90)(150,0)
\put(0,0){\insertfig{10}{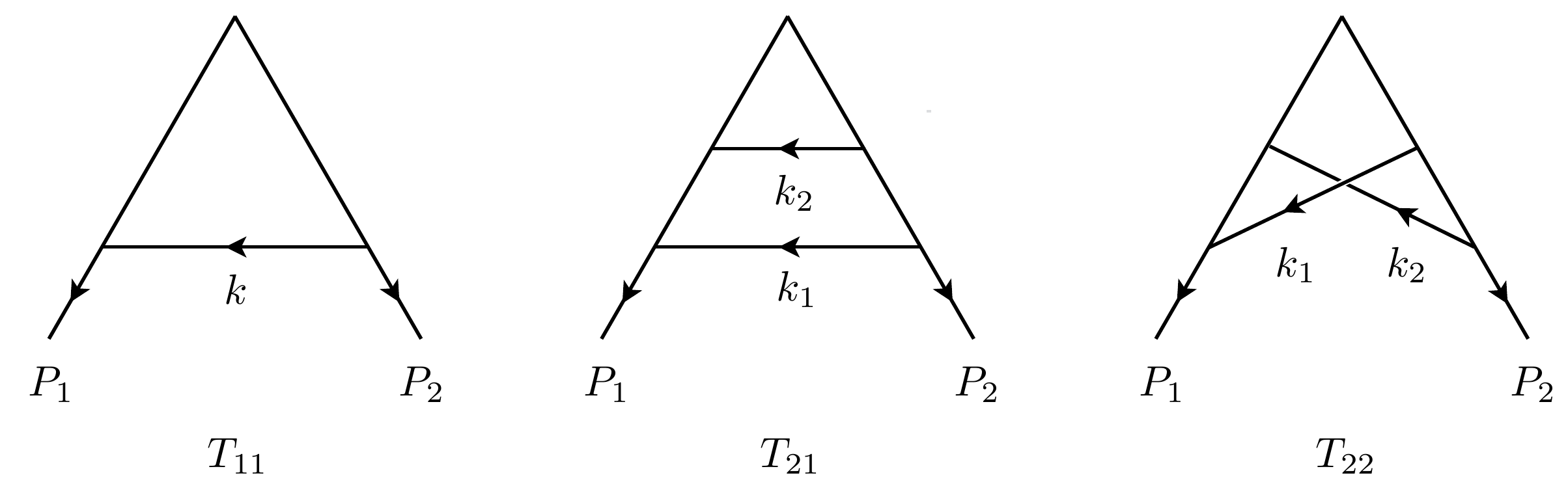}}
\end{picture}
}
\end{center}
\caption{\label{ScalarIntsFig} Scalar integrals defining the two-loop form factor.}
\end{figure}

The main object of our current analysis is the off-shell Sudakov form factor\footnote{Notice with our definition of amputated states, the
form factor $\mathcal{F}_2$ is dimensionless.} in the $\mathcal{N} = 4$ sYM
\begin{align}
\frac{1}{2 N_c}
\int d^4 x \, \e^{- i q \cdot x} \bra{P_1, P_2}  \tr_{\rm adj} \phi_{12}^2 (x) \ket{0}_{\rm amp}
=
(2 \pi)^4 \delta^{(4)} (q - P_1 - P_2) \mathcal{F}_2
\, ,
\end{align}
where $\phi = \phi_{12}$ of the sextet $\phi_{AB}$ of scalars of the $\mathcal{N} = 4$ sYM and, unless stated otherwise, all fields
are adjoint matrices $\phi = \phi_a T^a$ with the generators defined by the SU$(N_c)$ structure constants $(T^a)^{bc} = - i f^{abc}$.
The composite operator $\phi_{12}^2$ defining it is BPS and thus the matrix element on its left-hand side is ultraviolet finite. For finite 
virtualities $P_i^2 \neq 0$, it is infrared finite as well. It was claimed in the literature \cite{Caron-Huot:2021usw} that these virtualities 
can be generated in a gauge invariant manner by considering maximally supersymmetric theory in higher dimensions while restricting 
loop integrals to four dimensions only. Higher dimensional components of external momenta will play the role of virtualities for momenta 
$P_i$ \cite{Caron-Huot:2021usw}. This approach is equivalent to consideration of the $\mathcal{N} = 4$ sYM on the Coulomb branch \cite{HennGiggs1,Caron-Huot:2021usw}. Integrands of the Sudakov form factor are expected to be universal in all dimensions 
\cite{Gehrmann:2011xn}, so to obtain the form factor $\mathcal{F}_2$ on the Coulomb branch, i.e., the off-shell form factor we can 
use the naive off-shell continuation of on-shell result in $\mathcal{N} = 4$ sYM from the work of van Neerven from almost four decades ago 
\cite{vanNeerven:1985ja} (see also more recent \cite{Bork:2010wf,Gehrmann:2011xn}). According to this analysis, up to two-loop order, 
the perturbative expansion of $\mathcal{F}_2$ can be expressed in terms of just one and two scalar integrals at corresponding loop 
orders after Passarino-Veltman tensor reduction
\begin{align}
\label{2loopSudakov}
\mathcal{F}_2
= 1 + g^2 [- 2 T_{11} ] +  g^4 [  4 T_{21} + T_{22} ] + O (g^6)
\, .
\end{align}
They are the (iterated) ladders $T_{11}$, $T_{21}$ and the cross ladder $T_{22}$, as shown in Fig.\ \ref{ScalarIntsFig},
\begin{align}
\label{T11}
T_{11}
&
=
q^2 \int_k D (k) D (k - P_1) D (k + P_2)
\, , \\
\label{T21}
T_{21}
&
=
q^4
\int_{k_1, k_2} D(k_1) D(k_2) D(k_1 - P_1) D (k_1 + k_2 - P_1) D (k_1 + P_2) D (k_1 + k_2 + P_2)
\, , \\
\label{T22}
T_{22}
&
=
q^4
\int_{k_1, k_2} D(k_1) D(k_2) D(k_1 - P_1) D (k_1 + k_2 - P_1) D (k_2 + P_2) D (k_1 + k_2 + P_2)
\, ,
\end{align}
built from the products of the scalar propagators $D (k) = [k^2 + i 0]^{-1}$. The original analysis in Ref.\ \cite{vanNeerven:1985ja} 
was performed for massless on-shell legs $P_i^2 = 0$ \cite{Jackiw:1968zz} as distinguished it from its off-shell counterpart in the 
original Sudakov's treatment \cite{Sudakov:1954sw}, one needs to tame infrared divergences. A version of dimensional regularization
known as the supersymmetry-preserving dimensional reduction \cite{Siegel:1979wq} was employed there \cite{vanNeerven:1985ja}. 
So the above perturbative series is cast in terms of the $D$-dimensional 't Hooft coupling (with $g_{\rm\scriptscriptstyle YM}$ being the 
dimensionless bare Yang-Mills coupling constant)
\begin{align}
\label{DtHooft}
g^2 \equiv \e^{-\ep \gamma_{\rm\scriptscriptstyle E}} \frac{g_{\rm\scriptscriptstyle YM}^2 N_c}{(4 \pi)^{D/2}}
\, ,
\end{align}
and the $D$-dimensional loop-momentum integrations are performed with the $\overline{\rm MS}$-measure
\begin{align}
\int_k \equiv 
\e^{\ep \gamma_{\rm\scriptscriptstyle E}} \mu^{2 \ep} \int \frac{d^D k}{i \pi^{D/2}}
\, ,
\end{align}
where $D \equiv 4 - 2 \ep$. In spite of the fact that in our analysis the external lines are taken off the mass shell and thus the infrared 
singularities are regularized by $P_i^2 \neq 0$, we will adopt the same $D$-dimensional notations as given above since factorization 
of these finite integrals in terms of individual momentum regions inevitably and unavoidably induces divergences that need proper 
regularization. These will be of different nature depending on the assumed sign of the $\ep$ parameter, either ultraviolet or infrared 
as we will see below. But all of them will be unequivocally regularized by going to $D \neq 4$ dimensions.

In our earlier studies \cite{Belitsky:2022itf,Belitsky:2023ssv}, we calculated the off-shell Sudakov form factor up to three-loop order, i.e., 
an order of perturbation theory higher than displayed in Eq.\ \re{2loopSudakov}. We found that, for Euclidean external kinematics\footnote{By 
extracting the overall mass dimension of the form factor in terms of $Q^2$, $\mathcal{F}_2$ becomes a function of the ratio $m^2/Q^2$ only. 
So without loss of generality, we will set $Q^2 = 1$ from now on. This implies that all scalar integrals $T_{ij}$ are functions of the small variable
$m$ only, $T_{ij} = T_{ij} (m)$ (and, away from $D=4$, of $\mu$ measured in units of $Q$).}, in the near mass-shell limit
\begin{align}
q^2 = - Q^2 < 0
\, , \qquad
P_i^2 = - m^2 < 0
\, , \qquad
m^2/Q^2 \ll 1
\, ,
\end{align}
it exponentiates as shown in Eq.\ \re{ExactOffSud} including finite terms. 
Here, we expanded the exact results to $O(g^6)$,
\begin{align}
\label{OctAD2loops}
\Gamma_{\rm oct} (g) = 4 g^2 - 16 \zeta_2 g^4 + \dots 
\, , \qquad
D (g) = 4 \zeta_2 g^2 - 32 \zeta_4 g^4 + \dots
\end{align}
In the present study, we will dissect this result and stitch its various momentum components back together \` a 
la factorization theorem. 

Our considerations in Refs.\ \cite{Belitsky:2022itf,Belitsky:2023ssv} were based on two independent formalisms. One was grounded in a 
rigorous technique of deriving a Fuchsian system of linear differential equations \cite{Henn:2013pwa} for a set of the so-called Master 
Integrals (MIs) and solving the former interactively order by order in $\ep$. The scalar integrals $T_{ij}$ defining the form factor were then 
reduced to these MIs by employing integration-by-parts identities. The asymptotic limit of the near on-shell limit was taken only at the very 
end of the calculation. Another technique that we employed was far less robust. It carries on the status of experimental mathematics. It is 
known under the name of the Method of Regions (MofR) \cite{Beneke:1997zp}. It consists of determining all non-overlapping regions of 
loop-momentum integrations scaling appropriately with external kinematical limits imposed on observables in question, Taylor expanding 
integrands of Feynman integrals according to these and finally integrating every one of them over the {\sl entire} momentum space, dropping 
scaleless integrals along the way. Using this technique we recovered the result of the differential equations, thus, putting MofR on a firmer 
foundation for the problem at hand. It was shown that overlap contributions usually yield scaleless integrals which can be set to zero if they 
are regulated appropriately \cite{Jantzen:2011nz}.

In the present study, the MofR takes center stage!

\section{MofR assisted factorization}
\label{MofRfactor}

Traditionally factorization of multiscale Green's functions/matrix elements in terms of incoherent components responsible for physics 
at different momentum scales is accomplished by means of perturbative analysis of Feynman graphs' integrands making use of Landau 
equations \cite{Landau:1959fi,Sterman:1978bi,Collins11}. The latter are most straightforwardly formulated by employing Feynman 
parametrization of momentum integrals. Since this will be our main technical tool, it only makes more sense to introduce it early. 

Namely, for the $L_{ij}$-loop integrals in Eqs.\ \re{T11}-\re{T22}, we define
\begin{align}
\label{TijDef}
T_{ij} (m) = \int_0^\infty d^{N_{ij}} \bit{x}  \, \mathcal{J}_{N_{ij}, L_{ij}} \left(\bit{x}; m\right)
\, ,
\end{align}
with the integration measure and the integrand given by
\begin{align}
d^{N} \bit{x} 
=
\prod_{i = 1}^N d x_i \, \delta \left( \sum_{i=1}^N x_i - 1 \right)
\, , \quad
\mathcal{J}_{N,L} \left( \bit{x}; m \right) 
= 
\int_{k_1, \dots, k_L}
\frac{\Gamma (N)}{\left[- \sum_{i = 1}^N x_i D^{-1}_i \right]^{N}}
\, ,
\label{JfeynmanIntegrand}
\end{align}
respectively. A parameter $x_i$ is associated with each (inverse) Feynman propagator $D^{-1}$ so that $N_{ij}$ is their total number 
in a given graph.

Making use of these definitions, infrared singularities of Feynman integrals are determined by the Landau equations
\begin{align}
x_i D^{-1}_i = 0
\, , \qquad
\partial_{k_\ell} \sum_{i = 1}^N x_i D^{-1}_i = 0
\, .
\end{align}
The first of them states that infrared singularities of Feynman integrals stem either when each propagator in a graph goes on-shell or 
when the corresponding line is eliminated from it. The second condition implies that the integration contours cannot be deformed away 
from these singularities and thus they get {\sl pinched} by them. Every set of solutions to the Landau equations provides 
a {\sl pinched surface} where singularities reside and thus endow every Feynman graph with its reduced counterpart. However, our 
pinching will be a little bit more specific than traditional analyses \cite{Sterman:1978bi} in that every component of the reduced graph 
will have a precise operator definition involving {\sl unrestricted} integration over the entire momentum space. In this sense, our 
treatment is akin to the one performed with the framework of effective field theories \cite{Becher:2009qa,Feige:2014wja} or more recent 
approaches to on-shell physics \cite{Dixon:2008gr,Agarwal:2021ais}. However, the use of different regularization procedures will trickle 
down to differences in the form of factorized components involved.

\subsection{Geometric approach to MofR}

MofR formulated in the momentum space, as described at the end of Sect.\ \ref{OffShellSudakSect}, can be quite tedious when applied
to multiloop integrals \cite{Smirnov:1998vk} due to the necessity to cover the entire range of loop momenta with non-overlapping 
regions even though sometimes these generate scaleless integrals that vanish within dimensional regularization \cite{Jantzen:2011nz}.
Here is where the Feynman parametrization comes to the rescue again, as was demonstrated in Ref.\ \cite{Smirnov:1999bza}. It was 
shown there that all non-vanishing regions can be formulated in a covariant fashion independent of such nuisances as a choice of reference 
frame or loop-momentum routing within a graph. In addition, the Feynman integral representation (up to a stipulation to be pointed below) 
allows one for an entirely geometric way to unravel contributing regions \cite{Pak:2010pt}.

It proceeds as follows \cite{Pak:2010pt}, see also recent Refs.\ \cite{Heinrich:2021dbf,Gardi:2022khw}, for more comprehensive,
reader-friendly treatments. Integrating out the loop momenta in the Feynman integrand \re{JfeynmanIntegrand}, it is cast in terms 
of two Symanzik polynomials $U$ and $F$,
\begin{align}
\label{Jfeynman}
\mathcal{J}_{N,L} \left( \bit{x}; m \right) 
=
\e^{\ep \gamma_{\rm\scriptscriptstyle E} L} \mu^{2 \ep L} 
\Gamma \left(N - \ft12 L D\right) [ U(\bit{x}) ]^{N - \ft12 (L+1) D} [ F(\bit{x}; m) ]^{- N + \ft12 L D}
\, ,
\end{align}
with $U/F$ being independent of/dependent on the kinematical invariant $m$ and determined by the spanning trees ($T^1$)/two-trees ($T^2$), 
\begin{align}
U(\bit{x}) = \sum_{T^1} \prod_{i \not\in T^1} x_i
\, , \qquad
F(\bit{x}; m) = - \sum_{T^2} {\rm mom}^2_{T^2} \prod_{i \not\in T^2} x_i
\end{align}
respectively. They are homogeneous functions of $\bit{x}$ of degree $L$ and $L+1$, respectively, In the second formula, ${\rm mom}^2_{T^2}$ 
defines the squared sum of momenta entering one of the connectivity components of the two-tree \cite{Smirnov:2012gma}. In the present case, 
these are reduced to first-order polynomials in $m$ such that $F$ is given by the polynomial of the form
\begin{align}
F(\bit{x}; m) = \sum_{i = 1}^M c_i \, x_1^{r_1} \dots x_N^{r_N} m^{2 r_{N + 1}}
\, , 
\end{align}
with positive expansion coefficients $c_i$ and $r_i \in \{ 0, 1 \}$ . The same-sign nature of the $c_i$'s is the stipulation we alluded to earlier for 
the successful application of the geometric approach that we are turning to in the next paragraph. The same formula holds for $U$ except for a 
different set of integers $r_i \in \{ 0, 1 \}$ ($i = 1, \dots, N$) and with $r_{N+1} = 0$,
\begin{align}
U (\bit{x}) = \sum_{i = 1}^{\xbar{\scriptstyle M}} \, x_1^{r_1} \dots x_N^{r_N}
\, ,
\end{align} 
So the Symazik polynomials correspond to a set of $M$ points $\bit{r} = (r_1, \dots, r_N, r_{N+1})$ and $\xbar{M}$ points $\bar{\bit{r}} = 
(r_1, \dots, r_N, 0)$ in the $(N+1)$-dimensional vector space, respectively. However, the above homogeneity requirements further confine 
these points to the $N$-dimensional hyperplane $\sum_{i=1}^N r_i = L + 1$ for $F$ and $(N-1)$-dimensional hyperplane 
$\{\sum_{i=1}^N r_i = L, r_{N+1} = 0\}$ for $U$.

So in the Feynman parameter space, different scalings of the loop momenta in the small external parameter $m$ are traded for certain 
equivalent scalings of the integration variables $\bit{x}$. We substitute
\begin{align}
\bit{x} \to m^{2 \bit{\scriptstyle v}} \bit{x}
\, ,
\end{align}
with a $(N+1)$-dimensional vector $\bit{v}$ of integers, into Eq.\ \re{TijDef} with \re{Jfeynman} and after factoring the minimal powers
of $m$ from the Symanzik polynomials
\begin{align}
F(m^{2 \bit{\scriptstyle v}} \bit{x}; m) = m^{2 \min (\bit{\scriptstyle v} \cdot \bit{\scriptstyle r})} F_m (\bit{x})
\, , \quad
U(m^{2 \bit{\scriptstyle v}} \bit{x}) = m^{2 \min (\bit{\scriptstyle v} \cdot \bar{\bit{\scriptstyle r}})} U_m (\bit{x})
\, ,
\end{align}
the residual functions $F_m$ and $U_m$ admit a regular Taylor expansion in $m$. Then the leading asymptotic contribution
to the integral in question is determined by the first term in their Taylor expansion, i.e., 
\begin{align}
F(m^{2 \bit{\scriptstyle v}} \bit{x}; m) \to m^{2 \min (\bit{\scriptstyle v} \cdot \bit{\scriptstyle r})} F_0 (\bit{x})
\, , \quad
U(m^{2 \bit{\scriptstyle v}} \bit{x}) \to m^{2 \min (\bit{\scriptstyle v} \cdot \bar{\bit{\scriptstyle r}})} U_0 (\bit{x})
\, .
\end{align}
In this manner, the integral is reduced to an overall $\ep$-dependent power of $m$ accompanied by a coefficient given by
an $m$-independent integral of the residual Symanzik polynomials $ U_0 (\bit{x})$ and $F_0 (\bit{x})$.

How does one systematically choose the sought-after region vectors $\bit{v}$? An ingenious recipe was suggested in Ref.\ \cite{Pak:2010pt}.
To start with, in order to avoid dealing with two independent polynomials, consider the product 
\begin{align}
F(\bit{x}; m) U(\bit{x}) = \sum_{i = 1}^M c_i \, x_1^{r_1} \dots x_N^{r_N} m^{2 r_{N + 1}}
\, ,
\end{align}
with some constants $c_i$ (different from those for the individual Symanzik polynomial $F(\bit{x}; m)$) as well as the first $N$ components 
of the $(N+1)$-dimensional vectors $\bit{r} = (r_1, \dots, r_N, r_{N+1})$ now drawing values from a larger set $r_i \in \{0,1,2\}$. These
vectors define the vertices of a convex hull
\begin{align}
{\rm Newt} [UF] = \left\{ \sum_{i = 1}^M \alpha_i \bit{r}_i : \alpha_i \geq 0 \ \& \sum_{i=1}^M \alpha_i = 1 \right\}
\, .
\end{align}
This is the famed Newton polytope. Its co-dimension one faces are known as facets $F$ and their intersection provides a complementary
definition of ${\rm Newt} [UF]$. Let's construct all inward-pointing normal vectors $\bit{v}_f$ to $F$. A subset of these with a positive 
$(N+1)$-st component is a set of lower facets $F_+$. The region vectors $\bit{v}$ are identified by $\bit{v}_f$ with $f \in F_+$. This 
construction was implemented in the automatic code {\tt asy.m} \cite{Pak:2010pt,Jantzen:2012mw}, which is currently an integral part 
of {\tt FIESTA5} \cite{Smirnov:2021rhf}.

\subsection{One-loop: MofR and pinching}

Coming back to the Sudakov form factor, let us begin with the one-loop integral $T_{11}$. Assigning the Feynman parameters
to the scalar propagators as they appear in the integrand from left to right, the 4-dimensional Newton polytope possesses four
lower facets with the following region vectors\footnote{Here and below, we do not display the unit component along the $m$ axis.}
\begin{align}
\bit{v}_{\rm h} = (0,0,0)
\, , \qquad
\bit{v}_{\rm c1} = (0,0,1)
\, , \qquad
\bit{v}_{\rm c2} = (0,1,0)
\, , \qquad
\bit{v}_{\rm us} = (0,1,1)
\, .
\end{align}
The corresponding Feynman integrals and their solutions are
\begin{align}
T_{11}^{\rm h}
&
=
\e^{\ep \gamma_{\rm\scriptscriptstyle E}} \mu^{2 \ep} \Gamma (1 + \ep)
\int d^3 \bit{x} \, (x_1 + x_2 + x_3)^{-1 + 2 \ep} (x_2 x_3)^{-1 - \ep}
\nonumber\\
&\qquad\qquad\qquad\qquad
=
\e^{\ep \gamma_{\rm\scriptscriptstyle E}} \mu^{2 \ep} \frac{\Gamma^2 (- \ep) \Gamma (1 + \ep)}{\Gamma (1 - 2 \ep)}
\, , \\
\label{1loopT11c1}
T_{11}^{\rm c1}
&
=
\e^{\ep \gamma_{\rm\scriptscriptstyle E}} \mu^{2 \ep} \Gamma (1 + \ep)
\int d^3 \bit{x} \, (x_1 + x_2)^{-1 + 2 \ep} (x_1 x_2 + x_2 x_3)^{-1 - \ep}
\nonumber\\
&\qquad\qquad\qquad\qquad
=
\e^{\ep \gamma_{\rm\scriptscriptstyle E}} \left( \frac{\mu^2}{m^2} \right)^\ep \frac{\Gamma^2 (- \ep) \Gamma (\ep)}{2 \Gamma (- 2 \ep)}
\, , \\
\label{1loopT11c2}
T_{11}^{\rm c2}
&
=
\e^{\ep \gamma_{\rm\scriptscriptstyle E}} \mu^{2 \ep} \Gamma (1 + \ep)
\int d^3 \bit{x} \, (x_1 + x_3)^{-1 + 2 \ep} (x_1 x_3 + x_2 x_3)^{-1 - \ep}
\nonumber\\
&\qquad\qquad\qquad\qquad
=
\e^{\ep \gamma_{\rm\scriptscriptstyle E}} \left( \frac{\mu^2}{m^2} \right)^\ep \frac{\Gamma^2 (- \ep) \Gamma (\ep)}{2 \Gamma (- 2 \ep)}
\, , \\
T_{11}^{\rm us}
&
=
\e^{\ep \gamma_{\rm\scriptscriptstyle E}} \mu^{2 \ep}  \Gamma (1 + \ep)
\int d^3 \bit{x} \, x_1^{-1 + 2 \ep} (x_1 x_2 + x_2 x_3 + x_3 x_1)^{-1 - \ep}
\nonumber\\
&\qquad\qquad\qquad\qquad
=
\e^{\ep \gamma_{\rm\scriptscriptstyle E}} \left( \frac{\mu^2}{m^4} \right)^\ep \Gamma (1 - \ep) \Gamma (\ep)^2
\, .
\end{align}
Our next order of business is to figure out the reason behind the labeling we attributed to different regions. A naked-eye inspection 
demonstrates that the dependence on the soft scale $m$ is different for the three groups. The region vector $\bit{v}_{\rm h}$ does 
not yield any, while it gets singular (for $\ep > 0$) as we move to $\bit{v}_{{\rm c}i}$ becoming the strongest for $\bit{v}_{\rm us}$.

Let us extract the reduced graph of pinch regions for the $T_{11}$ integral from the above MofR consideration. To do it in a way that 
connects our analysis to traditional pinch-surface analyses, let us do it in the momentum space for the loop momentum $k$. As
exhibited by the components of the region vectors, the scaling of the Feynman parameters translates directly into the inverse 
scaling of the respective propagators
\begin{align}
x_i \sim m^{2 v_i} \qquad\to\qquad D_i^{- 2 v_i + n_i}
\, ,
\end{align}
up to an arbitrary additive overall shift $\bit{n} = (n, \dots, n, 0)$ with $n \in \mathbb{Z}_+$. So that shifting components in each
region by an integer to get proper scaling of the loop momentum with external scales
\begin{align}
\begin{array}{llll}
\bit{v}_{\rm h} = (0,0,0):     \quad & k^2 \sim m^0 \, , \quad & (k - P_1)^2 \sim m^0  \, , \quad & (k + P_2)^2 \sim m^0 , \\
\bit{v}_{\rm c1} = (-1,-1,0): \quad & k^2 \sim m^2 \, ,    \quad & (k - P_1)^2 \sim m^2  \, , \quad & (k + P_2)^2 \sim m^0 , \\
\bit{v}_{\rm c2} = (-1,0,-1): \quad & k^2 \sim m^2 \, ,    \quad & (k - P_1)^2 \sim m^0  \, , \quad & (k + P_2)^2 \sim m^2 , \\
\bit{v}_{\rm us} = (-2,-1,-1):\quad & k^2 \sim m^4 \, ,\quad & (k - P_1)^2 \sim m^2  \, , \quad & (k + P_2)^2 \sim m^2 ,
\end{array}
\end{align}
we conclude that $\bit{v}_{\rm h}$, $\bit{v}_{\rm c1}$, $\bit{v}_{\rm c2}$ and $\bit{v}_{\rm us}$ correspond to the hard\footnote{Recall 
that we have set the hard scale to one, $Q = 1$, to deal with just one dimensionless parameter $m$ in the problem. Otherwise, $k \sim Q$.}, 
$k \sim 1$; $P_1$-collinear, $k \sim P_1$; $P_2$-collinear, $k \sim P_2$ and ultrasoft, i.e., $k \sim m^2$, regions, respectively. So we have
three singular pinch surfaces for the one-loop integral: the two collinear regions, and the ultrasoft. As can be easily verified, they are 
given by the following loop-momentum integrals extended over the entire Minkowski domain
\begin{align}
T_{11}^{\rm c1}
&
=
- \int_k D (k) D (k - P_1) D_{\rm eik} (k + P_2)
\, , \\
T_{11}^{\rm c2}
&
=
- \int_k D (k) D_{\rm eik} (k - P_1) D (k + P_2)
\, , \\
\label{1loopUS}
T_{11}^{\rm us}
&
=
- \int_k D (k) D_{\rm us} (k - P_1) D_{\rm us} (k + P_2)
\, ,
\end{align}
where we introduced notations for the eikonal and ultrasoft propagators
\begin{align}
D_{\rm eik} (k_i \pm P_j) \equiv [\pm 2 k_i \cdot p_j ]^{-1} 
\, , \qquad
\label{USpropagator}
D_{\rm us} (k_i \pm p_j) \equiv [ P_j^2 \pm 2 k_i \cdot p_j ]^{-1} 
\, . 
\end{align}
Here we employed the Sudakov decomposition of the external off-shell momenta
\begin{align}
P_{1,2} \equiv p_{1,2} + \alpha_i \, p_{2,1} + \, p^\perp_{1,2}
\end{align}
in terms of a pair of light-like vectors $p_i$ ($i =1,2$), 
\begin{align}
p_i^2 = 0, \qquad p_i \cdot p^\perp_i = 0, \qquad p_1 \cdot p_2 = - \ft12
\, ,
\end{align}
such that $P_i$'s are predominantly directed along their low-case counterparts $p_i$'s. 

\begin{figure}[t]
\begin{center}
\mbox{
\begin{picture}(0,145)(70,0)
\put(0,0){\insertfig{5}{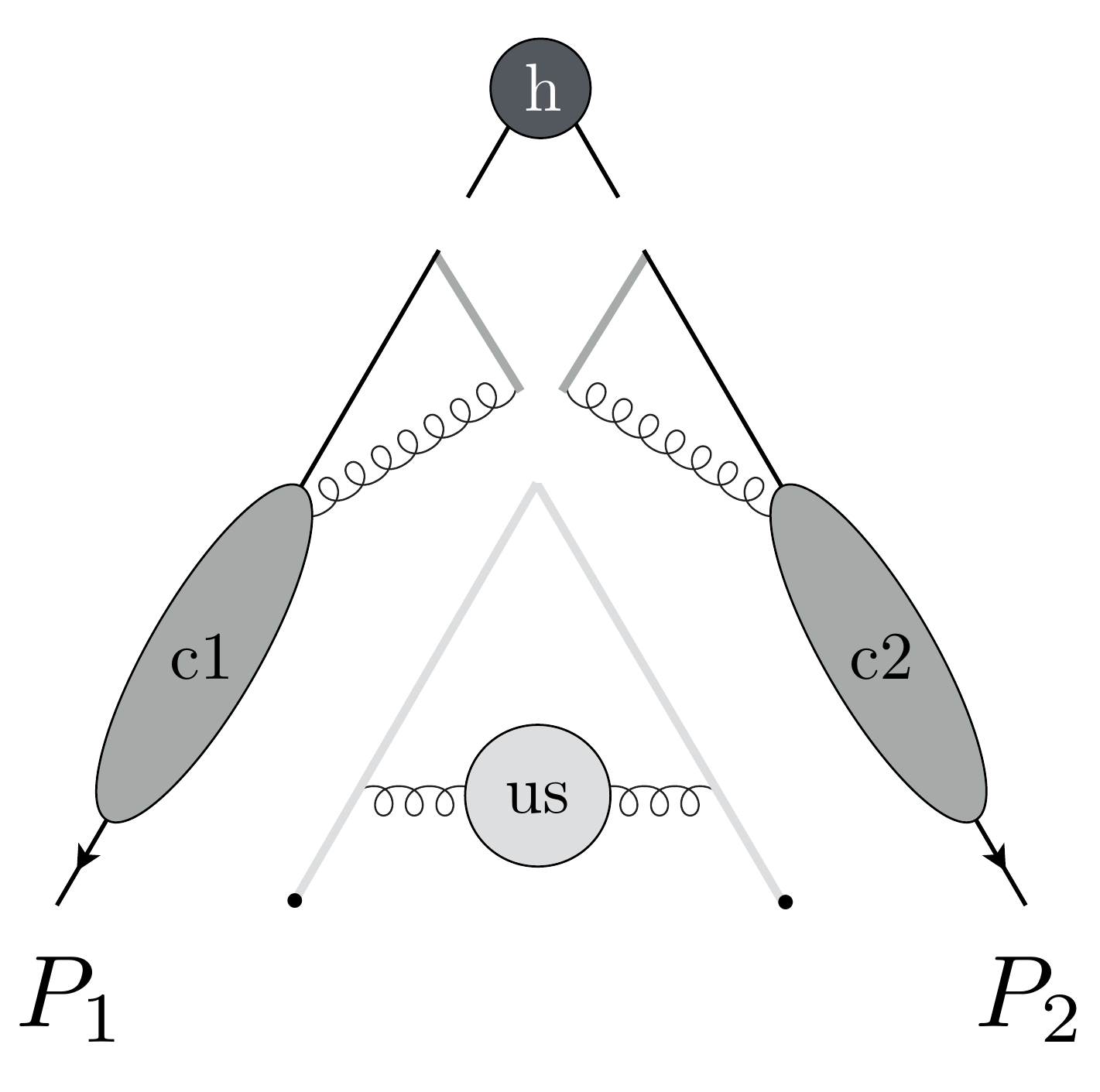}}
\end{picture}
}
\end{center}
\caption{\label{1loopPinchFig} Leading pinch regions of the one-loop integral $T_{11}$ for the off-shell Sudakov form factor.
The lighter the blob the softer are the loop momenta circulating in quantum loops.}
\end{figure}

The above integrals elucidate the underlying physical picture of the process. Namely, the two collinear regions describe the 
propagation of two energetic scalar jets with almost the speed of light. As a result, each one cannot resolve the internal content
of the other due to the Lorentz contraction and thus sees only its overall direction and color, but not more field-specific characteristics 
like spin. This is encoded in the eikonal form of the interaction. These jets can in turn interact with each other only through the ultrasoft 
gluon exchange, so as not to change the direction of their propagation. This is again encoded in the eikonal-like interactions
but they also accommodate for the non-vanishing virtuality of external lines. This is nothing else than the classical Coleman-Norton 
interpretation \cite{Coleman:1965xm} reflecting the pinch surfaces in the near mass-shell limit. Obviously, the hard region does 
not possess mass singularities as it is $m$-independent and is determined by the integral
\begin{align}
T_{11}^{\rm h}
&
=
- \int_k D_0 (k) D_0 (k - P_1) D_0 (k + P_2)
\, , 
\end{align}
with massless, i.e., on-shell external line, propagators $D_0 (k \pm P_i) = D (k \pm p_i)$.

To conclude, the one-loop analysis gives us the following asymptotic form of the one-loop graph as $m \to 0$
\begin{align}
T_{11} = T_{11}^{\rm h} + T_{11}^{\rm c1} + T_{11}^{\rm c2} + T_{11}^{\rm us}
\, ,
\end{align}
with the corresponding reduced one-loop graph on the leading pinch surface is shown in Fig. \ref{1loopPinchFig}. Let us point out that 
from the momentum-space perspective, an analogous one-loop analysis was performed in Ref. \cite{Becher:2014oda}.

\section{Jet function: definition and one-loop result}
\label{JetSect}

Having established the form of integrals for incoherent momentum components defining the pinched Sudakov form factor at one loop, 
we are now in a position to find their operator content. The advantage of the MofR compared to more traditional methods based on Landau
equations is that it provides us with a precise form of integrands valid over the entire Minkowski space: there is no need to assume
any momentum cut-offs.

We first turn to the collinear regions. Since both of them are identical up to the exchange of the legs' labels, it suffices to discuss just one.
With external scalar lines being off the mass-shell, this requires an interpretation for the states $\ket{P_1, P_2}_{\rm amp} = 
\ket{P_1}_{\rm amp} \otimes \ket{P_2}_{\rm amp}$. We understand them as
\begin{align}
\ket{P_i}_{\rm amp} \equiv \widetilde \phi^\dagger (P_i) \ket{0}_{\rm amp}
= 
\widetilde \phi^\dagger (P_i) \ket{0} [i D (P_i)]^{- 1}
\, ,
\end{align}
where $\widetilde \phi^\dagger$ is the Fourier transform of the scalar field $\phi^\dagger$ in the Heisenberg representation. Now, since the 
interaction of this leg with the one in the opposite jet is eikonal, this implies that the latter can be replaced by a Wilson line stretching from the
source to infinity along the light-like direction $p_i$. Thus we can immediately borrow the well-known definition of the jet function from the
on-shell case \cite{Collins:1989bt} and define
\begin{align}
\label{OffShellJet}
J_1 \equiv \bra{0} [\infty, 0]_{p_2} \phi (0) \ket{P_1}_{\rm amp}
\, ,
\end{align}
where we introduced the light-like Wilson line
\begin{align}
[\infty, 0]_{p} \equiv P \exp \left( i g_{\rm\scriptscriptstyle YM} \int_{0}^{\infty} d \tau \, p \cdot A (\tau p) \right)
\, .
\end{align}
There exists a potential subtlety due to the overcounting of the ultrasoft region from the off-shell leg in this definition. As in the on-shell 
approach, it can be eliminated by introducing the concept of the eikonal jet \cite{Dixon:2008gr}, i.e., when the scalar line itself is replaced 
by the Wilson line.  Within the framework of effective field theories, these are known as zero-bin subtractions \cite{Manohar:2006nz,Idilbi:2007ff}.
For our observable, the external leg is off the mass-shell and thus its treatment needs care. It will be performed in the next section.
The upshot is that the eikonal jet gets replaced for the case at hand by the ultrasoft jet function given by the vacuum expectation value
of the operator
\begin{align}
\label{USjet}
J^{\rm us}_1 \equiv 
\int_0^\infty d \sigma \, \e^{i \sigma P_1^2} \bra{0} [\infty, 0]_{p_2}  [2 p_1 \sigma , 0]^\dagger \ket{0}_{\rm amp}
\, .
\end{align}
The double counting is then removed by means of the substitution
\begin{align}
J_1 \to J_1/J^{\rm us}_1
\, .
\end{align}
The same holds for the other jet upon the interchange of the labels $1 \leftrightarrow 2$. Notice, however, since MofR universally employs
dimensional regularization to tame singularities, the ultrasoft jet function is given by scaleless integrals at each order of perturbation theory
and is thus set to one, $J^{\rm us}_{\rm dimreg} = 1$. It is a well-known fact in the effective theories as well \cite{Feige:2014wja}.

A one-loop calculation in the Feynman gauge then gives for the off-shell jet function \re{OffShellJet},
\begin{align}
J_1
= 1 
&
+ i g^2 \int_k \int_0^\infty d \sigma \, \e^{i \sigma (p_1 \cdot k)} p_1 \cdot (2 P_2 - k) D(k) D(k - P_2)
+
O(g^4)
\nonumber\\
&
=
1
-
g^2
\e^{\ep \gamma_{\rm\scriptscriptstyle E}} \left( \frac{\mu^2}{m^2} \right)^\ep \frac{\Gamma^2 (- \ep) \Gamma (\ep)}{2 \Gamma (- 2 \ep)}
+
O(g^4)
\, ,
\end{align}
and demonstrates the equivalence to MofR's $T_{11}^{\rm c1}$. A two-loop calculation of this jet factor will be performed elsewhere.

As a next step, let us focus on the consideration of the near mass-shell limit of the scalar Green function which was instrumental 
for the proper definition of the ultrasoft jet function introduced above. Along the way, we will uncover the form for an operator
defining the ultrasoft region of the pinched Sudakov form factor as well.

\section{Ultrasoft function}

Let us now discuss the ultrasoft function. It involves propagators in the near mass-shell limit. So we start our analysis by
considering their infrared properties.

\subsection{Definition}
\label{SectUSdef}

\begin{figure}[t]
\begin{center}
\mbox{
\begin{picture}(0,95)(230,0)
\put(0,0){\insertfig{16}{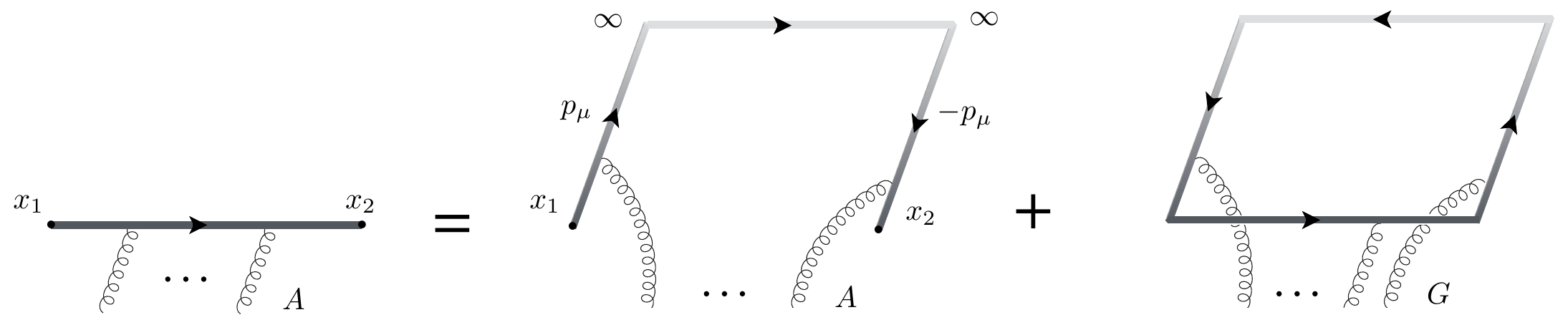}}
\end{picture}
}
\end{center}
\caption{\label{PropagatorFig} Leading singularity of the gauge-field propagator in the external field and contour deformation.}
\end{figure}

To analyze the infrared singularities of the scalar propagator, we will use a variant of the background field method \cite{Abbott:1980hw}.
It will echo the classical analysis of the infrared asymptotics of the electron Green function in QED \cite{Popov:1984mx}.

As was suggested by the analysis in the previous sections, ultrasoft singularities in Green functions stem from modes of very long 
wavelength. Thus, let us separate quantum gauge fields in terms of a sum of the fast $a$ and slow $A$ fields without specifying the 
precise location of the interface since it will not be relevant at the end, $A \to a + A$. Then, according to the analysis in the Appendix 
\ref{AppendixA}, the scalar propagator in the external field $A$ is given by the product
\begin{align}
D_A (x_2,x_1) = D_0 (x_2 -x_1) [x_2,x_1]
\, , 
\end{align}
where the slow modes enter only through the Wilson line along a straight segment connecting the points $x_1$ and $x_2$
\begin{align}
\label{SegmentWL}
[x_2,x_1] \equiv P \exp \left( i g_{\rm\scriptscriptstyle YM} \int_{x_1}^{x_2} dx \cdot A \right)
\, ,
\end{align}
and the hard field was integrated out. Without affecting the leading light-cone singularity of the scalar propagator, the integration contour 
entering the Wilson segment can be deformed to any other smooth shape since the difference is power-suppressed\footnote{See Ref.\
\cite{Falcioni:2019nxk} for a study relating anomalous dimension for various Wilson loop contours in the light-like case.}. For instance, as we 
demonstrated in Fig.\ \ref{PropagatorFig}, we can re-write $[x_2,x_1] $ identically as a difference between the staple-like contour and a 
closed loop. The gluons emitted by the closed loop are twist-one by default since they are expressible in terms of the gluon field strength
 tensor. Thus they yield power-suppressed contributions in the light-cone kinematics. Thus, up to power corrections the straight contour is 
 given the staple\footnote{The top of the staple can be ignored in all covariant gauges, but not in the light-like and axial types due 
 to zero modes which can propagate infinite distances, see, e.g., \cite{Belitsky:2002sm}.},
\begin{align}
[x_2,x_1] = [\infty, x_2]^\dagger_v [\infty, x_1]_v \big( 1 +  O(G_{\mu\nu}) \big)
\, ,
\end{align}
where the two rays are pointing along an arbitrary direction $v_\mu$. In particular, if $x_1 = 0$ and $x_2$ is changing along a
the same fixed vector $x_2 = \tau v$, $ [\infty, x_2]^\dagger_v [\infty, x_1]_v$ collapses on this line $[\tau v, 0]_v$.

For identification of the operator content of Eq. \re{1loopUS}, we need the Fourier transform of the scalar propagator. It reads
\begin{align}
i D_A (P) = \int d^D x \e^{i P \cdot x} D_0 (x) [x,0]
=
\int_0^\infty d \sigma \int \frac{d^D k}{(2 \pi)^D} \e^{i \sigma (P+k)^2} \widetilde{W} [k]
\, ,
\end{align}
here we used the Schwinger representation of the bare propagator $D (P+k)$ and the Fourier image of the Wilson segment
$$
\widetilde{W} [k] \equiv \int d^D x \e^{i x \cdot k}  [x,0]
\, .
$$ 
So far all of the transformations were exact. Now, for gluons connecting to the Wilson line in the ultrasoft region, i.e., $|P^2| \gg |k^2|$, 
we can expand the Lorentz-invariant exponent as $(P + k)^2 \simeq P^2 + 2 k \cdot p$ with light-like $p$, and perform the inverse Fourier 
transformation of $\widetilde{W} [k]$ back to the coordinate space. Finally, we deduce
\begin{align}
\label{PropExtFieldLC}
D_A (P) = 
\int_0^\infty d \sigma \e^{i \sigma P^2}  [2 \sigma p, 0]_p
\, ,
\end{align}
where we chose the link to be along $p$. As we can see from here, the nonzero virtuality of the scalar propagator introduces 
a segment Wilson line along the light-like direction $p_\mu$ and its finite extent is inversely proportional to the virtuality $P^2$.

Using Eq.\ \re{PropExtFieldLC}, we can immediately determine the operator definition of the ultrasoft factor,
\begin{align}
\label{IRfunction}
W (P_1, P_2) = \int_0^\infty d \sigma_1 d \sigma_2 \e^{i \sigma_1 P_1^2 + i \sigma_2 P_2^2} \mathcal{W} (\sigma_1, \sigma_2)
\, ,
\end{align}
with the integrand being
\begin{align}
\label{Wintegrand}
\mathcal{W} (\sigma_1, \sigma_2)
\equiv
\bra{0} [2 \sigma_1 p_1,0] [2 \sigma_2 p_2,0]^\dagger \ket{0}
\, .
\end{align}
Everywhere here and below it is tacitly implied that $P_i^2$ acquires a small positive imaginary part $P_i^2 + i 0_+$ for convergence
of the integrals in Eq.\ \re{IRfunction}. This quantity coincides with the so-called IR factor introduced several decades ago in Refs.\
\cite{Korchemsky:1985ts,Korchemsky:1988hd}. But this is it as far as our agreement goes. Our conclusions regarding its renormalization 
properties will differ.

Obviously, as it stands, the above vacuum expectation value \re{Wintegrand}, being defined by Wilson lines on finite intervals, is not 
gauge invariant: even a small gauge transformation of the gluon field will induce nonvanishing gauge matrices at its ends. However,
we are not actually after the ultrasoft function $W  (P_1, P_2) $ for finite virtualities but rather only for their asymptotically small 
values, i.e., on the pinch surface $P_i^2 \to 0$. The saddle-point approximation for \re{IRfunction} then implies that $\sigma_i$'s tend 
to infinity and, in this fashion, the integrand $\mathcal{W}$ of the ultrasoft function restores its gauge invariance. From the momentum-space 
perspective, this indicates that we are interested only in contributions at the poles $P_i^2 = 0$. Thus, it makes sense in what follows
to introduce an amputated ultrasoft function $S$ as a pole part of $W$
\begin{align}
\label{AmputatedUS}
W  (P_1, P_2) \equiv  [i D (P_1)] S (P_1, P_2) [i D (P_2)] + \mbox{less singular terms}
\, ,
\end{align}
with the scalar propagators $D (P_i) = [P_i^2]^{- 1}$, where, as alluded to earlier, we do not exhibit the imaginary shift. The residual
dependence of the left-hand side on $P_i^2$ is not rational and, as we will see below, is in fact non-analytic in these virtualities.

\begin{figure}[t]
\begin{center}
\mbox{
\begin{picture}(0,60)(230,0)
\put(0,0){\insertfig{16}{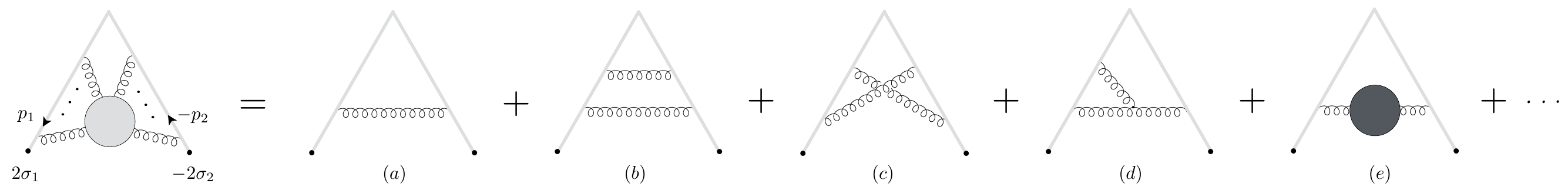}}
\end{picture}
}
\end{center}
\caption{\label{1and2loopIRFig} Graphs contributing to the one- $(a)$ and two-loop $(b-e)$ integrands $\mathcal{W}$ of the ultrasoft factor $W$.
A mirror image graph to $(d)$ is not shown.}
\end{figure}

\subsection{Perturbative expansion: One loop}
\label{US1looSect}

The ultrasoft function $W$ develops perturbative expansion in $g$. We will now calculate them order by order in the Feynman gauge. 
The light-like nature of the segments eliminates self-contractions of gluons from each of them individually. So we need to address only 
diagrams where the left link is connected to the right one. Up to two-loop order, its integrand receives contributions from Feynman 
diagrams shown in Fig.\ \ref{1and2loopIRFig}.

We start from the first quantum correction displayed in Fig.\  \ref{1and2loopIRFig} $(a)$. The integrand of the ultrasoft function reads
\begin{align}
{\mathcal W}^{(1)}_{(a)}
=
g_{\rm\scriptscriptstyle YM}^2 \int_0^{2 \sigma_1} d \tau_1 \int_{- 2 \sigma_2}^0 d \tau_2 
\vev{A_+ (p_1 \tau_1) A_- (- p_2 \tau_2)}_0 
\, ,
\end{align}
where $A_+ \equiv p_1 \cdot A$ and $A_- \equiv p_2 \cdot A$. Here $\vev{\dots}_0$ stands for the vacuum expectation value in the
interaction picture. Making use of the definition for the gluon propagator in the Feynman gauge
\begin{align}
\vev{A_\mu^a (x_1) A_\nu^b (x_2)}_0 \equiv - i \delta^{ab} g_{\mu \nu} D(x_1 - x_2)
\, , \qquad
\end{align}
and performing the half-range Fourier transform, its one-loop momentum-space integral representation becomes for the amputated
function \re{AmputatedUS}
\begin{align}
\label{1loopGluonExchange}
S^{(1)}_{(a)}
=
2 g^2 \int_k D (k) D_{\rm us} (k - p_1) D_{\rm us} (k + p_2)
\, .
\end{align}
It is written in terms of the momentum-space gluon propagator (denoted by the same letter as the coordinate one) 
\begin{align}
D(x) \equiv \int \frac{d^D k}{(2 \pi)^D} \e^{- i x \cdot k} D(k)
\, ,
\end{align}
as well as the ultrasoft ones \re{USpropagator}. The calculation of the resulting momentum integral is straightforward and we find for it
\begin{align}
S^{(1)}_{(a)}
=
- 2 g^2 w_{(a)} 
\, ,
\end{align}
where
\begin{align}
w_{(a)} 
= \e^{\ep \gamma_{\rm E}} \left( \frac{\mu^2}{m^4} \right)^\ep \Gamma^2 (\ep) \Gamma (1 - \ep)
\, .
\end{align}

\subsection{Two loops}
\label{US2looSect}

Let us now move on to the calculation of contributing two-loop graphs.

\subsubsection{Ladder graphs}

\begin{figure}[t]
\begin{center}
\mbox{
\begin{picture}(0,55)(115,0)
\put(0,0){\insertfig{8}{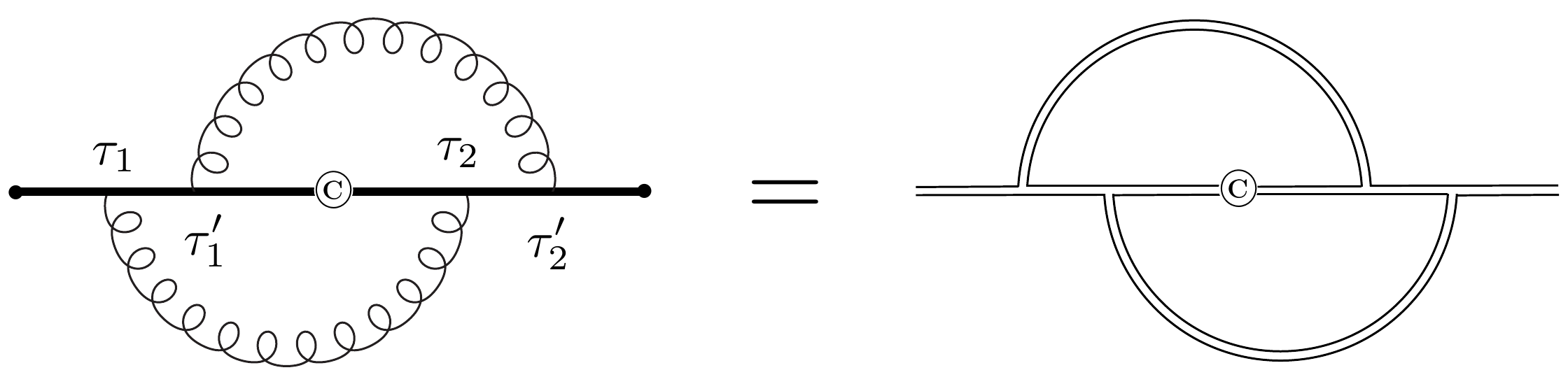}}
\end{picture}
}
\end{center}
\caption{\label{DoubleLineFig} A double line representation for the nonplanar ladder in Fig.\ \ref{1and2loopIRFig} $(c)$. The vertex
\copyright\ stands for the location of the cusp.}
\end{figure}

At two loops, we start with the iterated (planar and non-planar) ladders. Expanding each path-ordered exponents to $O(g^2)$, 
the integrals defining these contributions read
\begin{align}
{\mathcal W}^{(2)}_{(b+c)}
=
g_{\rm\scriptscriptstyle YM}^4 \int_0^{2 \sigma_1} d \tau_1 \int_0^{\tau_1} d \tau'_1 
\int_{- 2 \sigma_2}^0 d \tau_2 \int_{-2 \sigma_2}^{\tau_2} d \tau'_2 
\vev{A_+ (p_1 \tau_1) A_+ (p_1 \tau'_1) A_- (- p_2 \tau_2) A_- (- p_2 \tau'_2)}_0 
\, .
\end{align}
Wick contraction yields then planar and nonplanar double ladders
\begin{align}
&
\vev{A_+ (p_1 \tau_1) A_+ (p_1 \tau'_1) A_- (- p_2 \tau_2) A_- (- p_2 \tau'_2)}_0 
= \ft14 N_c^2
\\
&\qquad\qquad\qquad\qquad\times
\left[
D (p_1 \tau_1 + p_2 \tau'_2) D (p_1 \tau'_1 + p_2 \tau_2)
+
\ft12
D (p_1 \tau_1 + p_2 \tau_2) D (p_1 \tau'_1 + p_2 \tau'_2)
\right]
\, . \nonumber
\end{align}
As it is obvious from the double-line representation (see Fig.\ \ref{DoubleLineFig}), the nonplanar contribution is nevertheless of 
the leading color. Performing the half-range Fourier transformation, we find
\begin{align}
S^{(2)}_{(b+c)}
=
4 g^4 \left[ w_{(b)} + \ft12 w_{(c)} \right] 
\, ,
\end{align}
with
\begin{align}
w_{(b)} 
&
=
\int_{k_1, k_2} D(k_1) D(k_2) D_{\rm us} (k_1 - p_1) D_{\rm us} (k_1 + k_2 - p_1) D_{\rm us} (k_1 + p_2) D_{\rm us} (k_1 + k_2 + p_2)
\nonumber\\
&\qquad\qquad\qquad\qquad\qquad\qquad\qquad\qquad\qquad\qquad
=
\e^{2 \ep \gamma_{\rm E}} \left( \frac{\mu^2}{m^4} \right)^{2\ep} \Gamma^2 (- \ep) \Gamma^2 (2 \ep)
\, , \\
w_{(c)} 
&
=
\int_{k_1, k_2} D(k_1) D(k_2) D_{\rm us} (k_1 - p_1) D_{\rm us} (k_1 + k_2 - p_1) D_{\rm us} (k_2 + p_2) D_{\rm us} (k_1 + k_2 + p_2)
\nonumber\\
&\qquad\qquad\qquad\qquad\qquad\qquad\qquad\qquad\qquad\qquad
=
\e^{2 \ep \gamma_{\rm E}} \left( \frac{\mu^2}{m^4} \right)^{2 \ep} \Gamma^2 (- \ep) \Gamma^2 (2 \ep)
\, .
\end{align}
Both of them are given by identical expressions.

Notice that we have not used the non-Abelian exponentiation theorem in our treatment of iterated ladders. The reason being
that one has to perform the half-line Fourier transform at the end of the calculation which requires expanding the exponentiated
contribution in the perturbative series anyway.

\subsubsection{Non-Abelian graphs}

The contribution from the non-Abelian graph in Fig.\  \ref{1and2loopIRFig} $(d)$ to the integrand is
\begin{align}
{\mathcal W}^{(2)}_{(d)}
=
g_{\rm\scriptscriptstyle YM}^4 \int_0^{2 \sigma_1} d \tau_1 \int_0^{\tau_1} d \tau'_1 
\int_{- 2 \sigma_2}^0 d \tau_2
\int d^D x_0
\vev{
A_+ (2 p_1 \tau_1) A_+ (2 p_1 \tau'_1) A_- (- 2 p_2 \tau_2 ) V_{\rm g} (x_0)
}_0
\, ,
\end{align}
where out of three terms of the momentum-space Feynman rule for the three-gluon vertex $V_{\rm g} = 
f^{abc} (\partial_\mu A_\nu^a)  A_\mu^b  A_\nu^c$ only two contribute non-trivially giving
\begin{align}
{\mathcal W}^{(2)}_{(d)}
&=
\ft{i}{4} g^4 \int_0^{2 \sigma_1} d \tau_1 \int_0^{\tau_1} d \tau'_1 \int_{- 2 \sigma_2}^0 d \tau_2
\\
&\times
\int_{k_1, k_2}
\e^{- i  p_1 \cdot ( \tau_1 k_1 +  \tau'_1 k_2 ) + i \tau_2 \, p_2 \cdot (k_1 + k_2)}
p_1 \cdot (k_2 - k_1) D(k_1) D (k_2) D(k_1 + k_2)
\, . \nonumber
\end{align}
Let us describe the next couple of steps in more detail since this will exhibit the gauge-restoring limit we advocated at the end 
of Sect.\ \ref{SectUSdef}. Namely, performing the nested line integrals, we find
\begin{align}
{\mathcal W}^{(2)}_{(d)}
&=
\ft{i}{4} g^4 \int_{k_1, k_2} D(k_1) D (k_2) D(k_1 + k_2) 
E_{\sigma_2} (0; p_2 \cdot (k_1 + k_2))
\\
&\times
\big[
E_{\sigma_1} (0; p_1 \cdot k_1)
-
2 E_{\sigma_1} (0; p_1 \cdot (k_1 + k_2))
+
E_{\sigma_1} (p_1 \cdot k_1; p_1 \cdot (k_1 + k_2))
\big]\, . \nonumber
\end{align}
where we used the notation for the function
\begin{align}
E_{\sigma} (r_1; r_2) \equiv \frac{\e^{- 2 i \sigma r_1} - \e^{- 2 i \sigma r_2} }{r_1 - r_2}
\, .
\end{align}
In the infinite-segment limit, the first two terms in the square brackets generate nonvanishing contributions, while the last 
one vanishes due to the near-total cancellation of rapid oscillations. This can also be verified by performing the half-rage 
Fourier transforms and noticing that this term does not induce a pole in $P_1^2$. Neglecting it, we then find the contribution 
to the amputated ultrasoft function
\begin{align}
S^{(2)}_{(d)} = g^4 w_{(d)}
\, ,
\end{align}
with two-loop integral evaluated to
\begin{align}
w_{(d)}
&
= \int_{k_1, k_2} D(k_1) D (k_2) D(k_1 + k_2) D_{\rm us} (k_1 + k_2 + p_2)
\left[ 2 D_{\rm us} (k_1 + k_2 - p_1) - D_{\rm us} (k_1 - p_1)  \right]
\nonumber\\
&
= 2 \e^{2 \ep \gamma_{\rm E}} \left( \frac{\mu^2}{m^4} \right)^{2 \ep} \Gamma (- \ep) \Gamma (2 \ep)
\big[
\Gamma (- 2 \ep) \Gamma (2 \ep) \Gamma (1 + \ep)
-
\Gamma (1 - \ep) \Gamma (- 1 + 2 \ep)
\big]
\, .
\end{align}
Notice that the second term in the square bracket does not possess uniform transcendentality (UT) when expanded in $\ep$.
However, it has a bubble subgraph as its inner loop and it will be canceled, after doubling this contribution thanks to the 
mirror non-Abelian image to Fig.\ \ref{1and2loopIRFig} $(d)$, by the vacuum polarization diagram that we finally turn to next.

\subsubsection{Self-energy graphs}

The two-loop self-energy diagram in  Fig.\ \ref{1and2loopIRFig} $(e)$ is the only one at this loop order that explicitly depends
on the number of gluon polarization circulating in the loop, i.e., it displays full-fledged scheme dependence. To preserve supersymmetry
we have to choose Siegel's dimensional reduction \cite{Siegel:1979wq} as opposed to the conventional dimensional regularization.
To this end, we can borrow the result from Ref.\ \cite{Belitsky:2003ys}, where the one-loop effect from the $\mathcal{N}=4$ fields 
on the gluon propagator was calculated to all orders in $\ep$. Thus, we just need to substitute the gluon propagator $D(k)$ in
Eq.\ \re{1loopGluonExchange} with
\begin{align}
D (k) \to - 2 g^2 \left(\frac{\mu^2}{-k^2} \right)^\ep \e^{\ep \gamma_{\rm E}}
\frac{\Gamma (\ep) \Gamma^2 (1-\ep)}{\Gamma (2 - 2 \ep)}
\, .
\end{align}
The calculation of the remaining $k$-integrals gives for the amputated ultrasoft function
\begin{align}
S^{(2)}_{(e)} = g^4 w_{(e)}
\, ,
\end{align}
with 
\begin{align}
w_{(e)}
=
- 2 \left(\frac{\mu^2}{m^4} \right)^{2 \ep} 
\e^{2 \ep \gamma_{\rm E}}
\frac{\Gamma (\ep) \Gamma^2 (1-\ep)}{\Gamma (2 - 2 \ep)}
\frac{\Gamma (1 - 2 \ep) \Gamma^2 (2 \ep)}{\Gamma (1 + \ep)}
\, .
\end{align}
Elementary simplifications then demonstrate that indeed this is twice the UT-violating contribution from the non-Abelian diagram.
Thus this cancellation enforces the well-known UT feature of the maximally supersymmetric theory exhibited here through the
ultrasoft function.

\subsubsection{The sum}

Before we take the sum of all the graphs, we need to take care of the overall ultraviolet subtraction of the segmented Wilson line. 
However, the only one that would be needed for its infinite-$\sigma$ limit is the gauge invariant counterterm due to the coupling 
renormalization, i.e., proportional to the one-loop beta-function. The latter vanishes, however, and thus there are no further contributions 
to account for. Adding all of the above diagrams gives us the final expression
\begin{align}
S = 1 + g^2 [ - 2 T_{11}^{\rm us} ] + g^4 [ 4 T_{21}^{\rm us} + T_{22}^{\rm us} ] + O (g^6)
\, , 
\end{align}
with
\begin{align}
T_{11}^{\rm us} 
&
= \left(\frac{\mu^2}{m^4} \right)^{\ep} \e^{\ep \gamma_{\rm E}} \Gamma^2 (\ep) \Gamma (1 - \ep)
\, , \\
\label{T21WL}
T_{21}^{\rm us} 
&
= \left(\frac{\mu^2}{m^4} \right)^{2 \ep} \e^{2 \ep \gamma_{\rm E}} \Gamma^2 (2 \ep) \Gamma^2 (- \ep)
\, , \\
\label{T22WL}
T_{22}^{\rm us} 
&
= 2 \left(\frac{\mu^2}{m^4} \right)^{2 \ep} \e^{2 \ep \gamma_{\rm E}} \Gamma^2 (2 \ep) \Gamma (- \ep)
\big[
\Gamma (- \ep) -  \Gamma (1 - 2 \ep)  \Gamma (\ep)
\big]
\, ,
\end{align}
where we intentionally wrote it in a form that resembles the scalar integral expansion of the two-loop form factor. In
this manner, it will be easier for us to compare these expressions with the results of MofR to be performed in the following section. 
In fact, ultrasoft function exponentiates including
the finite terms and reads
\begin{align}
\label{UntwistedUS}
\log S 
=
- g^2 \left(\frac{\mu^2}{m^4} \right)^{\ep} 
\left( \frac{2}{\ep^2} + 3 \zeta_2 \right)
+
g^4 \left(\frac{\mu^2}{m^4} \right)^{2 \ep} 
\left( \frac{5 \zeta_2}{\ep^2} - \frac{7 \zeta_3}{\ep} +  71 \zeta_4 \right)
+
\dots
\, ,
\end{align}
with the exponent possessing at most order $\ep^{-2}$ poles to this order in coupling. This behavior persists to all orders in 't Hooft coupling
according to the conjecture we put forward in Ref.\ \cite{Belitsky:2022itf}.

\section{All-order factorization: a proposal}
\label{AllOrderFactSect}

With perturbative results obtained so far, we can now propose a naive factorization formula for the off-shell Sudakov form factor.
Namely, restoring the ultrasoft jet factors, anticipating a potential use of regularizations different from dimensional, we write for
the near mass-shell limit as
\begin{align}
\label{NaiveFactorization}
\mathcal{F}_2^{\rm naive} = 
H (\mu^2, \ep) 
\frac{J_1 (\mu^2/m^2, \ep)}{J^{\rm us}_1 (\mu^2/m^2, \ep)} \frac{J_2 (\mu^2/m^2, \ep)}{J^{\rm us}_2 (\mu^2/m^2, \ep)} 
S (\mu^2/m^4, \ep)
\, ,
\end{align}
where we assumed the same virtuality for both external legs, $P_i^2 = - m^2 \to 0$. The operator definitions for all pinch surfaces 
were introduced in Eqs.\ \re{OffShellJet}, \re{USjet}, \re{AmputatedUS}, respectively. $H$ is the hard matching coefficient. If one 
wishes to restore the dependence on the hard scale $Q^2$, the following substitutions are to be made
\begin{align}
\mu \to \mu/Q
\, , \quad
m \to m/Q
\, .
\end{align}
In this manner, one recognizes the two infrared scales involved in the problem, the collinear and the ultrasoft 
\cite{Fishbane:1971jz,Mueller:1981sg,Korchemsky:1988hd}
\begin{align}
\label{ScalesOffSud}
\mu_{\rm col} = m
\, , \qquad
\mu_{\rm us} = m^2/Q
\, .
\end{align}
Let us now proceed with its two-loop test.

\section{Two-loop test}
\label{MofR2loopTestSect}

As before, we will rely on MofR to verify the formula \re{NaiveFactorization}. Since we have a clear identification of the hard,
collinear and ultrasoft regions at one loop, which are associated with $\mu^{2 \ep}$, $(\mu^2/m^2)^\ep$ and $(\mu^2/m^4)^\ep$ 
scalings of contributing regions, respectively, it should be straightforward to do the same at two loops. With each loop integration 
producing each of the above, we anticipate six scalings:
\begin{align}
\begin{array}{lll}
\mbox{h-h}: \, \mu^{4 \ep}
\, , \quad
&
\mbox{h-c}: \, (\mu^4/m^2)^\ep
\, , \quad
&
\mbox{h-us}: \, (\mu^4/m^4)^\ep
\, , \\
\mbox{c-c}: \, (\mu^2/m^2)^{2\ep}
\, , \quad
&
\mbox{c-us}: \, (\mu^4/m^6)^\ep
\, , \quad
&
\mbox{us-us}: \, (\mu^2/m^4)^{2\ep}
\, .
\end{array}
\end{align}
The total number of regions in each two-loop graph can be up to sixteen via naive counting due to the doubling of the collinear ones, 
one for each leg, $\# ({\rm h}, {\rm c1}, {\rm c2}, {\rm us})^2 = 16$. However, there are far fewer in the iterated ladder since the inner, 
i.e., $k_2$-loop, is always harder than the outer one being closer to the source. In the crossed ladder case, on the other hand, both 
loop momenta enter symmetrically and all sixteen regions can and do realize. However, as we will see from our analysis, the precise 
identification of regions will be different from the above naive counting.  As we can see the double-collinear and hard-ultrasoft 
regions possess the same scaling. However, they are easy to disentangle from each other by studying an explicit form of components 
of the region vectors. 

\subsection{MofR vs.\ naive factorization}

Our analysis of associated Newton polytopes with {\tt asy} for the two scalar integrals $T_{21}$ and $T_{22}$, defined by Eqs. \re{T21} 
and \re{T22}, reveals 9 and 16 lower facets, respectively. We choose to label the regions in the order $k_2\text{-}k_1$ of the loop 
integrals reflecting the momentum flow in Fig.\ \ref{ScalarIntsFig}. They are defined by the normal vectors

\begin{figure}[t]
\begin{center}
\mbox{
\begin{picture}(0,70)(230,0)
\put(0,0){\insertfig{16}{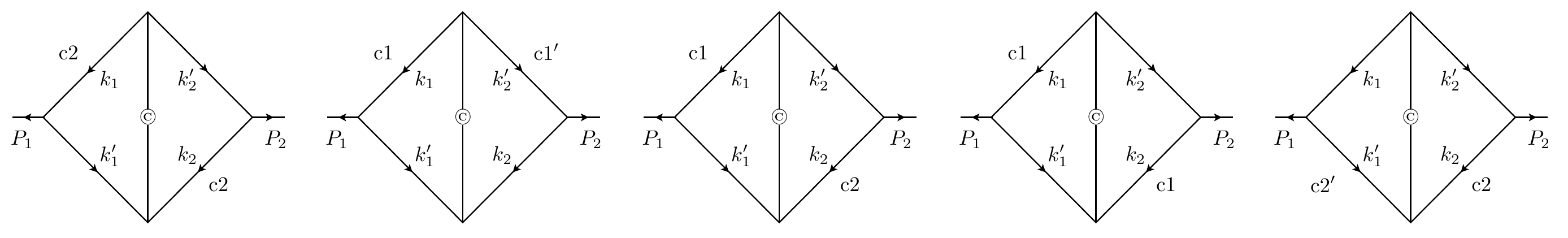}}
\end{picture}
}
\end{center}
\caption{\label{ColColFig} Momentum flow in the non-planar graph and all collinear-collinear regions in it.}
\end{figure}

\footnotesize
\begin{align}
\begin{array}{lll}
\bit{v}_{\rm h\text{-}h} = (0, 0, 0, 0, 0, 0)
\, , \ \
&
\bit{v}_{\rm h\text{-}c1} = (0, 1, 0, 1, 1, 1)
\, , \ \
&
\bit{v}_{\rm h\text{-}c2} = (0, 1, 1, 1, 0, 1)
\, , \\
\bit{v}_{\rm c1\text{-}c1} = (0, 0, 0, 0, 1, 1)
\, , \ \
&
\bit{v}_{\rm c2\text{-}c2} = (0, 0, 1, 1, 0, 0)
\, , \ \
&
\bit{v}_{\rm h\text{-}us} = (0, 2, 1, 2, 1, 2)
\, , \\
\bit{v}_{\rm c1\text{-}us} = (0, 1, 1, 1, 1, 2)
\, , \ \
&
\bit{v}_{\rm c2\text{-}us} = (0, 1, 1, 2, 1, 1)
\, , \ \
&
\bit{v}_{\rm us\text{-}us} = (0, 0, 1, 1, 1, 1)
\, ,
\end{array}
\end{align}
\normalsize

\noindent for $T_{21}$, and
\footnotesize
\begin{align}
\begin{array}{llll}
\bit{v}_{\rm h\text{-}h} 
= (0, 0, 0, 0, 0, 0)
\, , \ \
&
\bit{v}_{\rm c2\text{-}h} 
=
(1, 0, 1, 1, 0, 1) 
\, , \ \
&
\bit{v}_{\rm h\text{-}c1} 
=
(0, 1, 0, 1, 1, 1) 
\, , \ \
&
\bit{v}_{\rm c2\text{-}c2} 
=
(0, 0, 1, 1, 0, 0) 
\, , \\
\bit{v}_{\rm c1'\text{-}c1} 
=
(0, 1, 0, 1, 0, 0) 
\, , \ \
&
\bit{v}_{\rm c2\text{-}c1} 
=
(0, 0, 0, 1, 0, 1) 
\, , \ \
&
\bit{v}_{\rm c1\text{-}c1} 
=
(0, 0, 0, 0, 1, 1) 
\, , \ \
&
\bit{v}_{\rm c2\text{-}c2'} 
=
(1, 0, 0, 0, 0, 1) 
\, , \\
\bit{v}_{\rm us'\text{-}c1} 
=
(1, 1, 1, 2, 0, 1) 
\, , \ \
&
\bit{v}_{\rm c2\text{-}us} 
=
(0, 1, 1, 2, 1, 1) 
\, , \ \
&
\bit{v}_{\rm us\text{-}c1} 
=
(1, 0, 1, 1, 1, 2) 
\, , \ \
&
\bit{v}_{\rm c2\text{-}us'} 
=
(1, 1, 0, 1, 1, 2) 
\, , \\
\bit{v}_{\rm us'\text{-}us'} 
=
(1, 1, 0, 1, 0, 1) 
\, , \ \
&
\bit{v}_{\rm us\text{-}us} 
=
(0, 0, 1, 1, 1, 1) 
\, , \ \
&
\bit{v}_{\rm us'\text{-}us} 
=
(0, 1, 1, 2, 0, 0)
\, , \ \
&
\bit{v}_{\rm us\text{-}us'} 
=
(1, 0, 0, 0, 1, 2)
\, ,
\end{array}
\end{align}
\normalsize
for $T_{22}$.
What might be a bit puzzling is the number of collinear-collinear regions in the crossed ladder: there are five instead of the expected
four. But a quick look at the momentum flow in the graph, see, Fig.\ \ref{ColColFig}, confirms that this number is indeed correct. We 
also clarify in that figure the nomenclature used in the naming of the region vectors. Since we use non-symmetric routing of the loop 
momenta in an otherwise symmetric non-planar graph, the regions with primes in the $k_2\text{-}k_1$ labeling scheme correspond 
to the shifted momenta $k_2 \to k_2' = k_2 + p_2$ and $k_1 \to k_1' = k_1 - p_1$, respectively.

Feynman parameter integrals presented in Appendix \ref{AppendixB} were transformed into the Mellin-Barnes form making use
of the Mathematica package {\tt MBcreate.m} \cite{Belitsky:2022gba} and evaluated as a Laurent expansion in $\ep$. We 
found for these
\begin{align}
T_{21}^{\rm h\text{-}h}
&
=
\mu^{4 \ep}
\left[
\frac{1}{4 \ep^4} + \frac{5 \zeta_2}{4 \ep^2} + \frac{29 \zeta_3}{6 \ep} + \frac{135 \zeta_4}{16}
+ O (\ep)
\right]
\, \\
T_{21}^{\rm h\text{-}c1}
=
T_{21}^{\rm h\text{-}c2}
&
=
\left( \frac{\mu^4}{m^2} \right)^\ep 
\left[
- \frac{1}{2 \ep^4} - \frac{\zeta_2}{\ep^2} - \frac{11 \zeta_3}{3 \ep} - \frac{21 \zeta_4}{2}
+ O (\ep)
\right]
\, \\
\label{T21cici}
T_{21}^{\rm c1\text{-}c1}
=
T_{21}^{\rm c2\text{-}c2}
&
=
\left( \frac{\mu^2}{m^2} \right)^{2\ep}
\left[
\frac{1}{4 \ep^4} + \frac{\zeta_2}{4\ep^2} + \frac{5 \zeta_3}{6 \ep} + \frac{163 \zeta_4}{16}
+ O (\ep)
\right]
\, , \\
T_{21}^{\rm h\text{-}us}
&
=
\left( \frac{\mu^4}{m^4} \right)^{\ep}
\left[
\frac{1}{\ep^4} + \frac{\zeta_2}{\ep^2} - \frac{8 \zeta_3}{3 \ep} - \frac{5 \zeta_4}{4}
+ O (\ep)
\right]
\, , \\
T_{21}^{\rm c1\text{-}us}
=
T_{21}^{\rm c2\text{-}us}
&
=
\left( \frac{\mu^4}{m^6} \right)^{\ep} 
\left[
- \frac{1}{2 \ep^4} - \frac{\zeta_2}{\ep^2} + \frac{7 \zeta_3}{3 \ep} - 3 \zeta_4
+ O (\ep)
\right]
\, , \\
T_{21}^{\rm us\text{-}us}
&
=
\left( \frac{\mu^2}{m^4} \right)^{2 \ep}
\left[
\frac{1}{4 \ep^4} + \frac{5 \zeta_2}{4 \ep^2} - \frac{7 \zeta_3}{6 \ep} + \frac{159 \zeta_4}{16}
+ O (\ep)
\right]
\, .
\end{align}
and 
\begin{align}
T_{22}^{\rm h\text{-}h} 
&
=
\mu^{4 \ep}
\left[
\frac{1}{\ep^4} - \frac{6 \zeta_2}{\ep^2} - \frac{83 \zeta_3}{3 \ep} - \frac{177 \zeta_4}{4}
+ O (\ep)
\right]
\, , \\
T_{22}^{\rm c2\text{-}h} 
=
T_{22}^{\rm h\text{-}c1} 
&
=
\left( \frac{\mu^4}{m^2} \right)^{\ep}
\left[
-\frac{2}{\ep^4} + \frac{8 \zeta_2}{\ep^2} + \frac{100 \zeta_3}{3 \ep} + 63 \zeta_4
+ O (\ep)
\right]
\, , \\
\label{T22cici}
T_{22}^{\rm c2\text{-}c2} 
=
T_{22}^{\rm c1'\text{-}c1} 
=
T_{22}^{\rm c1\text{-}c1} 
=
T_{22}^{\rm c2\text{-}c2'} 
&
=
\left( \frac{\mu^2}{m^2} \right)^{2 \ep}
\left[
\frac{1}{2 \ep^4} - \frac{\zeta_2}{\ep^2} - \frac{53 \zeta_3}{6 \ep} - \frac{167 \zeta_4}{8}
+ O (\ep)
\right]
\, , \\
T_{22}^{\rm c2\text{-}c1} 
&
=
\left( \frac{\mu^2}{m^2} \right)^{2 \ep}
\left[
\frac{4}{\ep^4} - \frac{4 \zeta_2}{\ep^2} - \frac{56 \zeta_3}{3\ep} -21 \zeta_4
+ O (\ep)
\right]
\, , \\
T_{22}^{\rm us'\text{-}c1} 
=
T_{22}^{\rm c2\text{-}us} 
=
T_{22}^{\rm us\text{-}c1} 
=
T_{22}^{\rm c2\text{-}us'} 
&
=
\left( \frac{\mu^4}{m^6} \right)^{\ep}
\left[
-\frac{1}{\ep^4} - \frac{2 \zeta_2}{\ep^2} + \frac{14 \zeta_3}{3 \ep} -6 \zeta_4
+ O (\ep)
\right]
\, , \\
T_{22}^{\rm us'\text{-}us'} 
=
T_{22}^{\rm us\text{-}us} 
&
=
\left( \frac{\mu^2}{m^4} \right)^{2 \ep}
\left[
\frac{1}{4 \ep^4} + \frac{5 \zeta_2}{4 \ep^2} - \frac{7 \zeta_3}{6 \ep} + \frac{159 \zeta_4}{16}
+ O (\ep)
\right]
\, , \\
T_{22}^{\rm us'\text{-}us} 
=
T_{22}^{\rm us\text{-}us'} 
&
=
\left( \frac{\mu^2}{m^4} \right)^{2 \ep}
\left[
\frac{1}{4 \ep^4} + \frac{7 \zeta_2}{4 \ep^2} - \frac{2 \zeta_3}{3 \ep} + \frac{295 \zeta_4}{16}
+ O (\ep)
\right]
\, . 
\end{align}

With these findings in our hands, we can verify that the sum of double ultrasoft regions coincides with the Laurent expansion 
of the two-loop expressions $T_{21}^{\rm us}$ and $T_{22}^{\rm us}$ stemming from the Wilson line analysis, i.e., Eqs.\
\re{T21WL} and \re{T22WL},
\begin{align}
T_{22}^{\rm us} 
&= 
T_{21}^{\rm us\text{-}us}
+ O (\ep)
\, , \\
T_{22}^{\rm us} 
&= 
T_{22}^{\rm us'\text{-}us'} + T_{22}^{\rm us\text{-}us} + T_{22}^{\rm us'\text{-}us} + T_{22}^{\rm us\text{-}us'} 
+ O (\ep)
\, .
\end{align}
Therefore, it seems only natural to identify all double-collinear regions with labels ${{\rm c}i}$, ${{\rm c}i}'$ with the jet function $J_i$,
\begin{align}
J_i
=
1 + g^2 [ - 2 T_{11}^{{\rm c}i} ] + g^4 [ 4 T_{21}^{{\rm c}i} + T_{22}^{{\rm c}i} ] + O (g^6)
\, ,
\end{align}
where $T_{11}^{{\rm c}i}$ are given in Eq.\ \re{1loopT11c1} and \re{1loopT11c2} for $i=1,2$, respectively, while the two loop contributions are
\begin{align}
T_{21}^{\rm c1} = T_{21}^{\rm c1\text{-}c1}
\, , \quad
T_{21}^{\rm c2} = T_{21}^{\rm c2\text{-}c2}
\, , \quad
T_{22}^{\rm c1} = T_{22}^{\rm c1\text{-}c1} + T_{22}^{\rm c1'\text{-}c1} 
\, , \quad
T_{22}^{\rm c2} = T_{22}^{\rm c2\text{-}c2} + T_{22}^{\rm c2\text{-}c2'} 
\end{align}
with the right-hand sides given in Eqs.\ \re{T21cici} and \re{T22cici}. The region contribution $T_{22}^{\rm c2\text{-}c1}$ is not included 
since it is given by the product of the one-loop $T_{11}^{{\rm c}i}$'s, 
\begin{align}
\label{T22collinear}
T_{22}^{\rm c2\text{-}c1} 
=
4 T_{11}^{\rm c1} T_{11}^{\rm c2}
+ 
O (\ep)
\, ,
\end{align}
that emerges from the product $J_1 J_2$ of one-loop contributions. Finally, the two-loop hard matching coefficient is
\begin{align}
H = 1 + g^2 [ - 2 T_{11}^{\rm h} ] + g^4 [ 4 T_{21}^{\rm h\text{-}h} + T_{22}^{\rm h\text{-}h} ] + O (g^6)
\, .
\end{align}
This coincides with the massless result by van Neerven \cite{vanNeerven:1985ja}. All other mixed-region contributions should 
arise naturally in the product of factorized functions. However, this expectation turns out to be incorrect. There is a subtle mismatch 
between the two-loop ultra-soft--collinear regions and the product of one-loop ultrasoft and jet functions. Namely, the difference 
between Eqs.\ \re{2loopSudakov} and \re{NaiveFactorization} is
\begin{align}
\mathcal{F}_2 - \mathcal{F}_2^{\rm naive}
=
\left( Z^2 (\ep) - 1 \right) g^4 [-2 T_{11}^{\rm us}] [- 2 T_{11}^{\rm c1} - T_{11}^{\rm c2}]
\, ,
\end{align}
where 
\begin{align}
Z^2 (\ep) = \frac{\Gamma (1 + 2 \ep)}{\Gamma^2 (1 + \ep)}
=
1 + \zeta_2 \ep^2 + O (\ep^3)
\, .
\end{align}
Does this mean that our factorization formula \re{NaiveFactorization} is doomed? Not quite. 

\subsection{Twisted functions and finite renormalization}
\label{TwistSect}

Notice that $Z (\ep)$ is accompanied by the one-loop components of the factorization formula, so we can {\sl twist} them in such 
a manner that the desired factor of $Z^2$ arises in front of the product of ultrasoft and collinear functions, i.e., 
\begin{align}
T_{11}^{{\rm c}i} \to Z (\ep) T_{11}^{{\rm c}i}
\, , \quad 
T_{11}^{\rm us} \to Z (\ep) T_{11}^{\rm us}
\, ,
\end{align}
but it cannot affect the product of these with the hard function, so we have also to twist $T_{11}^{\rm h}$
\begin{align}
T_{11}^{\rm h} \to Z^{-1} (\ep) T_{11}^{\rm h}
\, .
\end{align}
But in these products, we have changed the finite parts of the one-loop functions involved (and their higher-order terms in the 
$\ep$-expansion which is not relevant however at this loop order), so we have to perform a transformation to neutralize this twist. 
The only parameter we have at our disposal is the 't Hooft coupling. The required finite scheme transformation that does the trick 
is
\begin{align}
g^2 \to g^2 \big(1 + 2 \zeta_2 g^2 \mu^{2 \ep}\big)
\, .
\end{align}
This reminds of a `physical' coupling that identifies the light-like cusp anomalous dimension as the new effective parameter of
perturbative expansion\footnote{We are grateful to Lorenzo Magnea for discussion of this point.} \cite{Catani:1990rr,Grozin:2015kna}. 
The twisted hard-matching coefficient and ultrasoft function are then
\begin{align}
\mathcal{H} 
&= 
(1 - 2 \zeta_2 g^2 \mu^{2 \ep})
\big[
1 + g^2 (1 + 2 \zeta_2 g^2 \mu^{2 \ep}) [ - 2 Z^{-1} T_{11}^{\rm h} ] + g^4 [ 4 T_{21}^{\rm h\text{-}h} + T_{22}^{\rm h\text{-}h} ] + O (g^6)
\big]
\, , \\
\mathcal{S} 
&= 
1 + g^2 (1 + 2 \zeta_2 g^2 \mu^{2 \ep})[ - 2 Z T_{11}^{\rm us} ] + g^4 [ 4 T_{21}^{\rm us} + T_{22}^{\rm us} ] + O (g^6)
\, ,
\end{align}
where in the former we also corrected for the change in the absolute normalization. The same holds for the
jet functions. For their product, we introduce 
\begin{align}
\mathcal{J}_1 \mathcal{J}_2
=
1 
&+ g^2 (1 + 2 \zeta_2 g^2 \mu^{2 \ep})[ - 2 Z T_{11}^{\rm c1} - 2 Z T_{11}^{\rm c2} ] 
\\
&
+ 
g^4 [ 4 T_{21}^{\rm c1} + T_{22}^{\rm c1} + 4 T_{21}^{\rm c2} + T_{22}^{\rm c2} + T_{22}^{\rm c2\text{-}c1} ] + O (g^6)
\, . \nonumber
\end{align}
Refactorization of their product into individual collinear components uses Eq.\ \re{T22collinear} and reads
\begin{align}
\mathcal{J}_1
=
1 + g^2 (1 + 2 \zeta_2 g^2 \mu^{2 \ep})[ - 2 Z T_{11}^{\rm c1} ] 
+ 
g^4 \big[ 4 T_{21}^{\rm c1} + T_{22}^{\rm c1} + 2 (1 - Z) [T_{11}^{\rm c1}]^2 \big] + O (g^6)
\, ,
\end{align}
and the same for $\mathcal{J}_2$ with the obvious replacement ${\rm c1} \to {\rm c2}$.

The final factorization formula for the off-shell Sudakov form factor then takes the same formal form as \re{NaiveFactorization},
where, however, one has to use the twisted form of individual functions with finite renormalization of the t Hooft coupling,
\begin{align}
\label{Factorization}
\mathcal{F}_2 = 
\mathcal{H} (\mu^2, \ep) \mathcal{J}_1 (\mu^2/m^2, \ep) \mathcal{J}_2 (\mu^2/m^2, \ep) \mathcal{S} (\mu^2/m^4, \ep)
\, .
\end{align}
Here, we do not display the ultrasoft jets since they are all unity in dimensional regularization employed in our formalism.

\section{Evolution equations}

Let us close this paper with a discussion of the renormalization and infrared evolution equations.

\subsection{Renormalization group}

While the total Sudakov form factor $\mathcal{F}_2$ is a finite quantity, its incoherent momentum components possess
divergences in the parameter of dimensional regularization $\ep$. They are infrared for $\mathcal{H}$, ultraviolet for 
$\mathcal{S}$, and mixed for the jet functions $\mathcal{J}_i$ canceling in their product. We can use the independence
of $\mathcal{F}_2$ on the renormalization scale $\mu$ to derive renormalization group equations for them. 

Let us define the renormalized functions in the minimal subtraction scheme as
\begin{align}
\mathcal{H} = \mathcal{Z}_{\rm h} \mathcal{H}_{\rm R} 
\, , \qquad
\mathcal{J}_i = \mathcal{Z}_{\rm col} \mathcal{J}_{i, \rm R}  
\, , \qquad
\mathcal{S} = \mathcal{Z}_{\rm us} \mathcal{S}_{\rm R}
\, ,
\end{align}
where to the two-loop order
\begin{align}
\log \mathcal{Z}_{\rm h} 
&
=
g^2
\left[ - \frac{2}{\ep^2} - \frac{2}{\ep} \log \mu^2 \right]
+
g^4
\left[ - \frac{\zeta_2}{\ep^2} - \frac{2 \zeta_2}{\ep} \log \mu^2 - \frac{3 \zeta_3}{\ep} \right]
\, , \\
\log \mathcal{Z}_{\rm col} 
&
=
g^2
\left[ \frac{2}{\ep^2} + \frac{2}{\ep} \log \mu^2 - \frac{2}{\ep} \log m^2 \right]
+
g^4
\left[ \frac{\zeta_2}{\ep^2} + \frac{2 \zeta_2}{\ep} \log \mu^2 + \frac{2 \zeta_2}{\ep} \log m^2 + \frac{3 \zeta_3}{\ep} \right]
\, , \\
\log \mathcal{Z}_{\rm us} 
&
=
g^2
\left[ - \frac{2}{\ep^2} - \frac{2}{\ep} \log \mu^2 + \frac{4}{\ep} \log m^2 \right]
+
g^4
\left[ - \frac{\zeta_2}{\ep^2} - \frac{2 \zeta_2}{\ep} \log \mu^2 - \frac{4 \zeta_2}{\ep} \log m^2 - \frac{3 \zeta_3}{\ep} \right]
\, .
\end{align}
After these subtractions, the logarithms of the renormalized functions admit the form
\begin{align}
\log \mathcal{H}_{\rm R} (\mu^2)
&
=
- g^2 \log^2 \mu^2 
+ 
g^4
\left[ - 2 \zeta_2 \log^2 \mu^2 - 6 \zeta_3 \log \mu^2 + \frac{9 \zeta_4}{2} \right]
\, , \\
\label{JetR}
\log \mathcal{J}_{i, \rm R} (\mu^2, m^2)
&
=
g^2 \left[ \log^2 \mu^2 - 2 \log \mu^2 \log m^2 +  \log^2 m^2 \right]
\\
&
+ 
g^4
\Big[ 2 \zeta_2 \log^2 \mu^2 + 4 \zeta_2 \log \mu^2 \log m^2 
- 4 \zeta_2 \log^2 m^2 
\nonumber\\
&\qquad\qquad\qquad\qquad\qquad\qquad\qquad\quad
+ 6 \zeta_3 \log \mu^2 - 6 \zeta_3 \log m^2 + \frac{\zeta_4}{2}  \Big]
\, , \nonumber\\
\label{USR}
\log \mathcal{S}_{\rm R} (\mu^2, m^2)
&
=
g^2 \left[ - \log^2 \mu^2 + 4 \log \mu^2 \log m^2 - 4 \log^2 m^2 - 4 \zeta_2 \right]
\\
&
+ 
g^4
\Big[ - 2 \zeta_2 \log^2 \mu^2 - 8 \zeta_2 \log \mu^2 \log m^2 + 16 \zeta_2 \log^2 m^2 
\nonumber\\
&\qquad\qquad\qquad\qquad\qquad\qquad\qquad\quad
- 6 \zeta_3 \log \mu^2 + 12 \zeta_3 \log m^2 + \frac{53 \zeta_4}{2}  \Big]
\, . \nonumber
\end{align}
Defining the anomalous dimensions with a conventional equation,
\begin{align}
\gamma = - \frac{d \log \mathcal{Z}}{d \log \mu^2}
\, ,
\end{align}
we immediately deduce their two-loop form
\begin{align}
\gamma_{\rm h} 
&
= g^2 \left[ - 2 \log \mu^2 \right]
+ g^4 \left[ - 4 \zeta_2 \log \mu^2 - 6 \zeta_3 \right]
\, , \\
\gamma_{\rm col} 
&
= g^2 \left[ 2 \log \mu^2 - 2 \log m^2 \right]
+ g^4 \left[  4 \zeta_2 \log \mu^2 + 4 \zeta_2 \log m^2 + 6 \zeta_3 \right]
\, , \\
\gamma_{\rm us} 
&
= g^2 \left[ - 2 \log \mu^2 + 4 \log m^2 \right]
+ g^4 \left[ - 4 \zeta_2 \log \mu^2 - 8 \zeta_2 \log m^2 - 6 \zeta_3 \right]
\, .
\end{align}
Here we used the fact that the 't Hooft coupling \re{DtHooft} is running in $D$-dimensions $d g/d\log\mu^2 = - \ep g/2$.
These anomalous dimensions are not independent and obey the equation
\begin{align}
\gamma_{\rm h} + 2 \gamma_{\rm col} + \gamma_{\rm us} = 0 
\end{align}
as a consequence of $\mu$-independence of $\mathcal{F}_2$. Notice that due to the fact that the ultrasoft
function is not scaleless, there is a mismatch between the infrared and ultraviolet logarithms. So contrary to
the massless case, where the study of ultraviolet renormalization properties of vacuum expectation values
of Wilson lines can be used to study infrared physics, this is no longer the case for the off-shell case.

\subsection{Infrared evolution}

Now we are in a position to derive evolution equations in the infrared scale $m$. In this section, we set $\mu = 1$,
i.e., $\mu$ set equal to the hard scale $Q$ in terms of the original dimensionful $\mu$. Making use of Eqs.\ \re{JetR} and \re{USR},
the dependence on $m$ is driven by the relations
\begin{align}
\label{JRirEvol}
\frac{d \log \mathcal{J}_{i, \rm R} (1, m^2)}{d \log m^2} 
&
= \frac{1}{2} \Gamma_{\rm oct} (g) \log m^2 + \Gamma_{\rm col} (g)
\, , \\
\label{SRirEvol}
\frac{d \log \mathcal{S}_{\rm R} (1, m^2)}{d \log m^2} 
&
= - 2 \Gamma_{\rm oct} (g)  \log m^2  - 2  \Gamma_{\rm col} (g)
\, ,
\end{align}
with the octagon anomalous dimension accompanying the logarithm \re{OctAD2loops} and the collinear anomalous dimension
for the log-free term,
\begin{align}
\Gamma_{\rm col} (g) = - 6 \zeta_3 g^4 + O (g^6)
\, .
\end{align}
It is not the on-shell collinear anomalous dimension, $\Gamma^{\rm on-shell}_{\rm col} (g) = - 4 \zeta_3 g^4 + O (g^6)$! The contribution 
from $\Gamma_{\rm col}$ cancels however in the evolution equation for the Sudakov form factor $\mathcal{F}_2$ such it does not possess
a single logarithmic term to any order of perturbation theory.

\subsection{Untwisted vs. twisted}

Finally, let us comment on the evolution equation for the untwisted ultrasoft functions $S$. It was proposed
in Ref.\ \cite{Korchemsky:1988hd} that its dependence on the ultrasoft scale $\mu_{\rm us}$ \re{ScalesOffSud}
is governed by the ubiquitous cusp anomalous dimension. We can now verify this statement using our explicit two-loop
result. We find from Eq.\ \re{UntwistedUS} for its renormalized counterpart ($\mu = 1$)
\begin{align}
\label{RenormalizedNaiveS}
\log S_{\rm R} (\mu_{\rm us})
=
g^2 \left[ - 4 \log^2 \mu_{\rm us}  - 3 \zeta_2 \right]
+
g^4 \left[ 40 \zeta_2 \log^2 \mu_{\rm us} + 28 \zeta_3 \log \mu_{\rm us} + 71 \zeta_4 \right]
+
O (g^6)
\, ,
\end{align}
and thus
\begin{align}
\frac{\log S_{\rm R} (\mu_{\rm us})}{d \log \mu_{\rm us}}
=
-
2 \left[ 4 g^2 - 40 \zeta_2 \right] \log \mu_{\rm us} - 28 \zeta_3 g^4 + O (g^6)
\, .
\end{align}
We observe that the function of the gauge coupling in front of $\log \mu_{\rm us}$ is not proportional to $\Gamma_{\rm cusp}$.
Let us point out, however, that if we modify the structure of the evolution equation by allowing a convolution term in its right-hand 
side, we can then recover the perturbative expansion of Eq.\ \re{RenormalizedNaiveS} with just the cusp anomalous dimension 
alone (and a new collinear one). Namely,
\begin{align}
\frac{d S_{\rm R} (\mu_{\rm us})}{d \log \mu_{\rm us}}
=
- [2 \Gamma_{\rm cusp} (g) \log \mu_{\rm us} + \widetilde\Gamma (g)] S_{\rm R} (\mu_{\rm us})
- 2  \Gamma_{\rm cusp} (g) \int_0^{\mu_{\rm us}} d \mu'_{\rm us} 
\frac{S_{\rm R} (\mu'_{\rm us}) - S_{\rm R} (\mu_{\rm us})}{\mu'_{\rm us} - \mu_{\rm us}}
\, ,
\end{align}
with $\Gamma_{\rm cusp} (g) = 4 g^2 - 8 \zeta_2 g^4 + O (g^6)$ and $\widetilde\Gamma (g) = - 92 \zeta_3 g^4 + O (g^6)$. The same can 
be done with Eqs.\ \re{JRirEvol}, \re{SRirEvol} but only making use of two different anomalous dimensions $\widetilde\Gamma (g)$, which 
is quite unnatural considering the fact that the log-free portion of the evolution equation should be absent for the total form factor.

\section{Conclusions}

In this paper, we continued our exploration of the Sudakov form factor on the Coulomb branch of $\mathcal{N} = 4$ sYM.
We analyzed its factorization in terms of incoherent components responsible for physics at different momentum scales,
from hard to ultrasoft. The main mathematical tool at our disposal was MofR which allowed us to identify various contributions 
to Feynman integrals in terms of essential momentum modes. However, when we attempted to naively separate these in
terms of the jet and ultrasoft functions, and a hard matching coefficient, we encountered a predicament at subleading orders
in the $\ep$-expansion: there was a `non-factorizable' leftover. We, however, managed to twist the functions involved along
with a finite renormalization of the 't Hooft coupling to rescue our factorization formula. The question remains whether
this can be done to all orders. It is nevertheless puzzling that while the complete form factor admits a very simple form in the 
near mass-shell limit, its factorization does not enjoy the simplicity of its on-shell counterpart, which is a far more complex function 
of the 't Hooft coupling, though not so different in kinematical dependence. Are we missing an integral convolution in disguise 
between different factorized components that would accommodate the deviation we observed in our two-loop analysis? This would, 
of course, be quite a departure from the familiar on-shell story. This question begs for further studies. Currently, this question is 
under study at three-loop order. 

\begin{acknowledgments}
We are grateful to Lorenzo Magnea for reading the manuscript and providing instructive comments. The work of A.B.\ was supported 
by the U.S.\ National Science Foundation under grant No.\ PHY-2207138. The work of L.B.\ was supported by the Foundation for the 
Advancement of Theoretical Physics and Mathematics “BASIS”. The work of V.S.\ was supported by the Russian Science Foundation 
under the agreement No.\ 21-71-30003.
\end{acknowledgments}

\appendix

\section{Light-cone singularity of scalar propagator}
\label{AppendixA}

As in the main body of the paper, we adopt notations from Ref.\ \cite{Belitsky:2003sh} for $\mathcal{N} = 4$ sYM Lagrangian 
and fields populating it. We consider the exact scalar propagator in the external long-wavelength gauge field $A$ and expand 
it to the linear order in $g$
\begin{align}
S^{ABCD}_{ab} (x_2, x_1) = \vev{\phi^{AB}_a (x_2) \phi^{CD}_b (x_1)}_A
= \delta_{ab} S^{ABCD}_{(0)} (x_2, x_1) + S^{ABCD}_{(1)ab} (x_2, x_1) + \dots
\, ,
\end{align}
where the leading term is the bare propagator
\begin{align}
S^{ABCD}_{(0)} (x_2, x_1) 
=
- \varepsilon_{ABCD} \frac{1}{4 \pi^{D/2}} \frac{\Gamma (D/2 - 1)}{[-x_{12}^2]^{D/2 - 1}}
\, ,
\end{align}
while the first correction takes the form of the vacuum expectation value with the gauge-scalar-scalar vertex $V_{\rm s}$
\begin{align}
S^{ABCD}_{(1)ab} (x_2, x_1)
&=
\vev{\phi^{AB}_a (x_2) \phi^{CD}_b (x_1) i \int d^D x_0 V_{\rm s} (x_0)}
\nonumber\\
&=
\frac{i g}{16 \pi^D} f_{abc} \varepsilon^{ABCD} (\partial_2 - \partial_1)_\mu
\int d^D x_0 A^a_\mu (x_0) \frac{\Gamma (D/2 - 1)}{[- x_{10}^2]^{D/2 - 1}} \frac{\Gamma (D/2 - 1)}{[- x_{20}^2]^{D/2 - 1}}
\, .
\end{align}
The integral can be easily calculated 
\begin{align}
&\int d^D x_0 A^a_\mu (x_0) \frac{\Gamma (D/2 - 1)}{[- x_{10}^2]^{D/2 - 1}} \frac{\Gamma (D/2 - 1)}{[- x_{20}^2]^{D/2 - 1}}
\nonumber\\
&\qquad\qquad\qquad=
\int_0^1 d\sigma (\sigma \bar\sigma)^{D/2 - 2} \int d^Dx_0 A^a_\mu (x_0 + \sigma x_2 + \bar{\sigma} x_1) \frac{\Gamma (D - 1)}{[- x_0^2 - \sigma \bar{\sigma} x_{12}^2]^{D - 2}}
\nonumber\\
&\qquad\qquad\qquad=
- i \pi^{D/2} \frac{\Gamma (D/2 - 1)}{[-x_{12}^2]^{D/2 - 2}} \int_0^1 d\sigma A_\mu^a (\sigma x_2 + \bar{\sigma} x_1)
\, ,
\end{align}
where $\bar\sigma \equiv 1 - \sigma$ and in the second step we performed the expansion of $A_\mu (x_0 + \sigma x_2 + \bar{\sigma} x_1)$ in 
the Taylor series around $x_0 = 0$ using the long wavelength approximation and eliminated higher-twist effects. In this manner, we find for
the leading light-cone singularity of $O(A)$ correction to the scalar propagator 
\begin{align}
S^{ABCD}_{(1)ab} (x_2, x_1)
= S^{ABCD}_{(0)} (x_2, x_1)
\left( i g  \int_0^1 d\sigma \, x_{12}^\mu (A_\mu)^{ab} (\sigma x_2 + \bar{\sigma} x_1) \right)
\, ,
\end{align}
with the adjoint matrix of gauge fields $(A_\mu)^{ab} = A_\mu^c (T^c)_{ab}$. Performing identical manipulations for terms of $O (A^2)$ and
higher, we uncover the known expression for the light-cone singularity of the propagator in the external field \cite{Gross:1971wn}
\begin{align}
S^{ABCD}_{(1)ab} (x_2, x_1) = S^{ABCD}_{(0)} (x_2, x_1) [x_2, x_1]
\, ,
\end{align}
with the Wilson line segment defined in Eq.\ \re{SegmentWL}.

\section{Region integrals at two loops}
\label{AppendixB}

In this appendix, we present contributions from separate regions to the two-loop ladder and cross-ladder integrals.
Their Feynman-parameter representations read for $T_{21}$

\footnotesize
\begin{align}
T_{21}^{\rm h\text{-}h}
&
=
\e^{2 \ep \gamma_{\rm\scriptscriptstyle E}} \mu^{4 \ep} \Gamma (2 + 2\ep)
\\
&\times
\int d^6 \bit{x} \, 
(x_1 x_2 + x_2 x_3 + x_1 x_4 + x_2 x_4 + x_3 x_4 + x_2 x_5 + x_4 x_5 + x_1 x_6 +  x_2 x_6 + x_3 x_6 + x_5 x_6)^{3 \ep} 
\nonumber\\
&\times
(x_2 x_3 x_5 + x_2 x_4 x_5 + x_3 x_4 x_5 + x_2 x_3 x_6 + x_1 x_4 x_6 + x_2 x_4 x_6 + x_3 x_4 x_6 + x_3 x_5 x_6 + x_4 x_5 x_6)^{-2 - 2 \ep}
\, , \nonumber\\
T_{21}^{\rm h\text{-}c1}
&
=
\e^{2 \ep \gamma_{\rm\scriptscriptstyle E}}  \left( \frac{\mu^4}{m^2} \right)^\ep \Gamma (2 + 2\ep)
\\
&\times
\int d^6 \bit{x} \, 
(x_1 x_2 + x_2 x_3 + x_1 x_4 + x_3 x_4 + x_1 x_6 + x_3 x_6)^{3 \ep} 
\nonumber\\
&\times
(x_1 x_2 x_3 + x_1 x_3 x_4 + x_2 x_3 x_5 + x_3 x_4 x_5 + x_1 x_3 x_6 +  x_2 x_3 x_6 + x_1 x_4 x_6 + x_3 x_4 x_6 + x_3 x_5 x_6)^{-2 - 2 \ep}
\, , \nonumber\\
T_{21}^{\rm h\text{-}c2}
&
=
\e^{2 \ep \gamma_{\rm\scriptscriptstyle E}} \left( \frac{\mu^4}{m^2} \right)^\ep \Gamma (2 + 2\ep)
\\
&\times
\int d^6 \bit{x} \, 
(x_1 x_2 + x_1 x_4 + x_2 x_5 + x_4 x_5 + x_1 x_6 + x_5 x_6)^{3 \ep} 
\nonumber\\
&\times
(x_1 x_2 x_5 + x_2 x_3 x_5 + x_1 x_4 x_5 + x_2 x_4 x_5 + x_3 x_4 x_5 + x_1 x_4 x_6 + x_1 x_5 x_6 + x_3 x_5 x_6 + x_4 x_5 x_6)^{-2 - 2 \ep}
\, , \nonumber\\
T_{21}^{\rm c1\text{-}c1}
&
=
\e^{2 \ep \gamma_{\rm\scriptscriptstyle E}} \left( \frac{\mu^2}{m^2} \right)^{2\ep} \Gamma (2 + 2\ep)
\\
&\times
\int d^6 \bit{x} \, 
(x_1 x_2 + x_2 x_3 + x_1 x_4 + x_2 x_4 + x_3 x_4)^{3 \ep}
\nonumber\\
&\times
(x_1 x_2 x_3 + x_1 x_2 x_4 + x_1 x_3 x_4 + x_2 x_3 x_5 + x_2 x_4 x_5 + x_3 x_4 x_5 + x_2 x_3 x_6 + x_1 x_4 x_6 + x_2 x_4 x_6 + x_3 x_4 x_6)^{-2 - 2 \ep}
\, , \nonumber\\
T_{21}^{\rm c2\text{-}c2}
&
=
\e^{2 \ep \gamma_{\rm\scriptscriptstyle E}}  \left( \frac{\mu^2}{m^2} \right)^{2\ep} \Gamma (2 + 2\ep)
\\
&\times
\int d^6 \bit{x} \, 
(x_1 x_2 + x_2 x_5 + x_1 x_6 + x_2 x_6 + x_5 x_6)^{3 \ep} 
\nonumber\\
&\times
(x_1 x_2 x_5 + x_2 x_3 x_5 + x_2 x_4 x_5 + x_1 x_2 x_6 + x_2 x_3 x_6 + x_1 x_4 x_6 + x_2 x_4 x_6 + x_1 x_5 x_6 + x_3 x_5 x_6 + x_4 x_5 x_6)^{-2 - 2 \ep}
\, , \nonumber\\
T_{21}^{\rm h\text{-}us}
&
=
\e^{2 \ep \gamma_{\rm\scriptscriptstyle E}}  \left( \frac{\mu^4}{m^4} \right)^{\ep} \Gamma (2 + 2\ep)
\\
&\times
\int d^6 \bit{x} \, 
(x_1 x_2 + x_1 x_4 + x_1 x_6)^{3 \ep} 
\nonumber\\
&\times
(x_1 x_2 x_3 + x_1 x_3 x_4 + x_1 x_2 x_5 + x_2 x_3 x_5 + x_1 x_4 x_5 + x_3 x_4 x_5 + x_1 x_3 x_6 + x_1 x_4 x_6 + x_1 x_5 x_6 + x_3 x_5 x_6)^{-2 - 2 \ep}
\, , \nonumber\\
T_{21}^{\rm c1\text{-}us}
&
=
\e^{2 \ep \gamma_{\rm\scriptscriptstyle E}}  \left( \frac{\mu^4}{m^6} \right)^{\ep} \Gamma (2 + 2\ep)
\\
&\times
\int d^6 \bit{x} \, 
(x_1 x_2 + x_1 x_4)^{3 \ep} 
\nonumber\\
&\times
(x_1 x_2 x_3 + x_1 x_2 x_4 + x_1 x_3 x_4 + x_1 x_2 x_5 + x_2 x_3 x_5 + x_1 x_4 x_5 + x_2 x_4 x_5 + x_3 x_4 x_5 + x_1 x_4 x_6)^{-2 - 2 \ep}
\, , \nonumber\\
T_{21}^{\rm c2\text{-}us}
&
=
\e^{2 \ep \gamma_{\rm\scriptscriptstyle E}} \left( \frac{\mu^4}{m^6} \right)^{\ep} \Gamma (2 + 2\ep)
\\
&\times
\int d^6 \bit{x} \, 
(x_1 x_2 + x_1 x_6)^{3 \ep} 
\nonumber\\
&\times
(x_1 x_2 x_3 + x_1 x_2 x_5 + x_2 x_3 x_5 + x_1 x_2 x_6 + x_1 x_3 x_6 + x_2 x_3 x_6 + x_1 x_4 x_6 + x_1 x_5 x_6 + x_3 x_5 x_6)^{-2 - 2 \ep}
\, , \nonumber\\
T_{21}^{\rm us\text{-}us}
&
=
\e^{2 \ep \gamma_{\rm\scriptscriptstyle E}} \left( \frac{\mu^2}{m^4} \right)^{2 \ep} \Gamma (2 + 2\ep)
\\
&\times
\int d^6 \bit{x} \, 
(x_1 x_2)^{3 \ep} 
\nonumber\\
&\times
(x_1 x_2 x_3 + x_1 x_2 x_4 + x_1 x_2 x_5 + x_2 x_3 x_5 + x_2 x_4 x_5 + x_1 x_2 x_6 + x_2 x_3 x_6 + x_1 x_4 x_6 + x_2 x_4 x_6)^{-2 - 2 \ep}
\, , \nonumber
\end{align}
\normalsize
and $T_{22}$
\footnotesize
\begin{align}
T_{22}^{\rm h\text{-}h} 
&
=
\e^{2 \ep \gamma_{\rm\scriptscriptstyle E}} \mu^{4 \ep} \Gamma (2 + 2\ep)
\\
&\times
\int d^6 \bit{x} \, 
(x_1 x_2 + x_2 x_3 + x_1 x_4 + x_2 x_4 + x_3 x_4 + x_1 x_5 + x_2 x_5 + x_3 x_5 + x_1 x_6 + x_3 x_6 + x_4 x_6 + x_5 x_6)^{3 \ep} 
\nonumber\\
&\times
(x_2 x_3 x_5 + x_1 x_4 x_5 + x_2 x_4 x_5 + x_3 x_4 x_5 + x_1 x_4 x_6 + x_4 x_5 x_6)^{-2 - 2 \ep}
\, , \nonumber\\
T_{22}^{\rm c2\text{-}h} 
&
=
\e^{2 \ep \gamma_{\rm\scriptscriptstyle E}} \left( \frac{\mu^4}{m^2} \right)^{\ep} \Gamma (2 + 2\ep)
\\
&\times
\int d^6 \bit{x} \, 
(x_1 x_2 + x_2 x_3 + x_2 x_4 + x_2 x_5 + x_1 x_6 + x_3 x_6 + x_4 x_6 + x_5 x_6)^{3 \ep} 
\nonumber\\
&\times
(x_2 x_3 x_5 + x_2 x_4 x_5 + x_1 x_2 x_6 + x_2 x_3 x_6 + x_1 x_4 x_6 +  x_2 x_4 x_6 + x_2 x_5 x_6 + x_4 x_5 x_6)^{-2 - 2 \ep}
\, , \nonumber\\
T_{22}^{\rm h\text{-}c1} 
&
=
\e^{2 \ep \gamma_{\rm\scriptscriptstyle E}} \left( \frac{\mu^4}{m^2} \right)^{\ep} \Gamma (2 + 2\ep)
\\
&\times
\int d^6 \bit{x} \, 
(x_1 x_2 + x_2 x_3 + x_1 x_4 + x_3 x_4 + x_1 x_5 + x_3 x_5 + x_1 x_6 + x_3 x_6)^{3 \ep}
 \nonumber\\
&\times
(x_1 x_2 x_3 + x_1 x_3 x_4 + x_1 x_3 x_5 + x_2 x_3 x_5 + x_1 x_4 x_5 + x_3 x_4 x_5 + x_1 x_3 x_6 + x_1 x_4 x_6)^{-2 - 2 \ep}
\, , \nonumber\\
T_{22}^{\rm c2\text{-}c2} 
&
=
\e^{2 \ep \gamma_{\rm\scriptscriptstyle E}} \left( \frac{\mu^2}{m^2} \right)^{2 \ep} \Gamma (2 + 2\ep)
\\
&\times
\int d^6 \bit{x} \, 
(x_1 x_2 + x_1 x_5 + x_2 x_5 + x_1 x_6 + x_5 x_6)^{3 \ep} 
\nonumber\\
&\times
(x_1 x_2 x_5 + x_2 x_3 x_5 + x_1 x_4 x_5 + x_2 x_4 x_5 + x_1 x_2 x_6 + x_1 x_4 x_6 + x_2 x_5 x_6 + x_4 x_5 x_6)^{-2 - 2 \ep}
\, , \nonumber\\
T_{22}^{\rm c1'\text{-}c1} 
&
=
\e^{2 \ep \gamma_{\rm\scriptscriptstyle E}} \left( \frac{\mu^2}{m^2} \right)^{2 \ep} \Gamma (2 + 2\ep)
\\
&\times
\int d^6 \bit{x} \, 
(x_1 x_5 + x_3 x_5 + x_1 x_6 + x_3 x_6 + x_5 x_6)^{3 \ep} 
\nonumber\\
&\times
(x_1 x_3 x_5 + x_2 x_3 x_5 + x_1 x_4 x_5 + x_3 x_4 x_5 + x_1 x_3 x_6 + x_1 x_4 x_6 + x_3 x_5 x_6 + x_4 x_5 x_6)^{-2 - 2 \ep}
\, , \nonumber\\
T_{22}^{\rm c2\text{-}c1} 
&
=
\e^{2 \ep \gamma_{\rm\scriptscriptstyle E}} \left( \frac{\mu^2}{m^2} \right)^{2 \ep} \Gamma (2 + 2\ep)
\\
&\times
\int d^6 \bit{x} \, 
(x_1 x_2 + x_2 x_3 + x_1 x_6 + x_3 x_6)^{3 \ep} 
\nonumber\\
&\times
(x_1 x_2 x_3 + x_2 x_3 x_5 + x_1 x_2 x_6 + x_1 x_3 x_6 + x_2 x_3 x_6 + x_1 x_4 x_6)^{-2 - 2 \ep}
\, , \nonumber\\
T_{22}^{\rm c1\text{-}c1} 
&
=
\e^{2 \ep \gamma_{\rm\scriptscriptstyle E}} \left( \frac{\mu^2}{m^2} \right)^{2 \ep} \Gamma (2 + 2\ep)
\\
&\times
\int d^6 \bit{x} \, 
(x_1 x_2 + x_2 x_3 + x_1 x_4 + x_2 x_4 + x_3 x_4)^{3 \ep}
\nonumber\\
&\times
(x_1 x_2 x_3 + x_1 x_2 x_4 + x_1 x_3 x_4 + x_2 x_3 x_5 + x_1 x_4 x_5 + x_2 x_4 x_5 + x_3 x_4 x_5 + x_1 x_4 x_6)^{-2 - 2 \ep}
\, , \nonumber\\
T_{22}^{\rm c2\text{-}c2'} 
&
=
\e^{2 \ep \gamma_{\rm\scriptscriptstyle E}} \left( \frac{\mu^2}{m^2} \right)^{2 \ep} \Gamma (2 + 2\ep)
\\
&\times
\int d^6 \bit{x} \, 
(x_2 x_3 + x_2 x_4 + x_3 x_4 + x_3 x_6 + x_4 x_6)^{3 \ep} 
\nonumber\\
&\times
(x_2 x_3 x_5 + x_2 x_4 x_5 + x_3 x_4 x_5 + x_2 x_3 x_6 + x_1 x_4 x_6 + x_2 x_4 x_6 + x_3 x_4 x_6 + x_4 x_5 x_6)^{-2 - 2 \ep}
\, , \nonumber\\
T_{22}^{\rm us'\text{-}c1} 
&
=
\e^{2 \ep \gamma_{\rm\scriptscriptstyle E}} \left( \frac{\mu^4}{m^6} \right)^{\ep} \Gamma (2 + 2\ep)
\\
&\times
\int d^6 \bit{x} \, 
(x_1 x_6 + x_3 x_6 + x_5 x_6)^{3 \ep} 
\nonumber\\
&\times
(x_2 x_3 x_5 + x_1 x_2 x_6 + x_1 x_3 x_6 + x_2 x_3 x_6 + x_1 x_4 x_6 + x_2 x_5 x_6 + x_3 x_5 x_6 + x_4 x_5 x_6)^{-2 - 2 \ep}
\, , \nonumber\\
T_{22}^{\rm c2\text{-}us} 
&
=
\e^{2 \ep \gamma_{\rm\scriptscriptstyle E}} \left( \frac{\mu^4}{m^6} \right)^{\ep} \Gamma (2 + 2\ep)
\\
&\times
\int d^6 \bit{x} \, 
(x_1 x_2 + x_1 x_5 + x_1 x_6)^{3 \ep} 
\nonumber\\
&\times
(x_1 x_2 x_3 + x_1 x_2 x_5 + x_1 x_3 x_5 + x_2 x_3 x_5 + x_1 x_4 x_5 + x_1 x_2 x_6 + x_1 x_3 x_6 + x_1 x_4 x_6)^{-2 - 2 \ep}
\, , \nonumber\\
T_{22}^{\rm us\text{-}c1} 
&
=
\e^{2 \ep \gamma_{\rm\scriptscriptstyle E}} \left( \frac{\mu^4}{m^6} \right)^{\ep} \Gamma (2 + 2\ep)
\\
&\times
\int d^6 \bit{x} \, 
(x_1 x_2 + x_2 x_3 + x_2 x_4)^{3 \ep} 
\nonumber\\
&\times
(x_1 x_2 x_3 + x_1 x_2 x_4 + x_2 x_3 x_5 + x_2 x_4 x_5 + x_1 x_2 x_6 + x_2 x_3 x_6 + x_1 x_4 x_6 + x_2 x_4 x_6)^{-2 - 2 \ep}
\, , \nonumber\\
T_{22}^{\rm c2\text{-}us'} 
&
=
\e^{2 \ep \gamma_{\rm\scriptscriptstyle E}} \left( \frac{\mu^4}{m^6} \right)^{\ep} \Gamma (2 + 2\ep)
\\
&\times
\int d^6 \bit{x} \, 
(x_2 x_3 + x_3 x_4 + x_3 x_6)^{3 ep} 
\nonumber\\
&\times
(x_1 x_2 x_3 + x_1 x_3 x_4 + x_2 x_3 x_5 + x_3 x_4 x_5 + x_1 x_3 x_6 + x_2 x_3 x_6 + x_1 x_4 x_6 + x_3 x_4 x_6)^{-2 - 2 \ep}
\, , \nonumber\\
T_{22}^{\rm us'\text{-}us'} 
&
=
\e^{2 \ep \gamma_{\rm\scriptscriptstyle E}} \left( \frac{\mu^2}{m^4} \right)^{2 \ep} \Gamma (2 + 2\ep)
\\
&\times
\int d^6 \bit{x} \, 
(x_3 x_6)^{3 ep} 
\nonumber\\
&\times
(x_2 x_3 x_5 + x_3 x_4 x_5 + x_1 x_3 x_6 + x_2 x_3 x_6 + x_1 x_4 x_6 +  x_3 x_4 x_6 + x_3 x_5 x_6 + x_4 x_5 x_6)^{-2 - 2 \ep}
\, , \nonumber\\
T_{22}^{\rm us\text{-}us} 
&
=
\e^{2 \ep \gamma_{\rm\scriptscriptstyle E}} \left( \frac{\mu^2}{m^4} \right)^{2 \ep} \Gamma (2 + 2\ep)
\\
&\times
\int d^6 \bit{x} \, 
(x_1 x_2)^{3 \ep} 
\nonumber\\
&\times
(x_1 x_2 x_3 + x_1 x_2 x_4 + x_1 x_2 x_5 + x_2 x_3 x_5 + x_1 x_4 x_5 + x_2 x_4 x_5 + x_1 x_2 x_6 + x_1 x_4 x_6)^{-2 - 2 \ep}
\, , \nonumber\\
T_{22}^{\rm us'\text{-}us} 
&
=
\e^{2 \ep \gamma_{\rm\scriptscriptstyle E}} \left( \frac{\mu^2}{m^4} \right)^{2 \ep} \Gamma (2 + 2\ep)
\\
&\times
\int d^6 \bit{x} \, 
(x_1 x_5 + x_1 x_6 + x_5 x_6)^{3 \ep} 
\nonumber\\
&\times
(x_1 x_2 x_5 + x_1 x_3 x_5 + x_2 x_3 x_5 + x_1 x_4 x_5 + x_1 x_2 x_6 + x_1 x_3 x_6 + x_1 x_4 x_6 + x_2 x_5 x_6 + x_3 x_5 x_6 + x_4 x_5 x_6)^{-2 - 2 \ep}
\, , \nonumber\\
T_{22}^{\rm us\text{-}us'} 
&
=
\e^{2 \ep \gamma_{\rm\scriptscriptstyle E}} \left( \frac{\mu^2}{m^4} \right)^{2 \ep} \Gamma (2 + 2\ep)
\\
&\times
\int d^6 \bit{x} \, 
(x_2 x_3 + x_2 x_4 + x_3 x_4)^{3 \ep} 
\nonumber\\
&\times
(x_1 x_2 x_3 + x_1 x_2 x_4 + x_1 x_3 x_4 + x_2 x_3 x_5 + x_2 x_4 x_5 + x_3 x_4 x_5 + x_2 x_3 x_6 + x_1 x_4 x_6 + x_2 x_4 x_6 + x_3 x_4 x_6)^{-2 - 2 \ep}
\, , \nonumber
\end{align}
\normalsize
respectively.


\end{document}